\documentclass[
 reprint,
 superscriptaddress,
 nofootinbib,
 amsmath,amssymb,
 aps,
 prl,
]{revtex4-2}

\usepackage{xcolor}
\usepackage{graphicx}
\usepackage{dcolumn}
\usepackage{bm}
\usepackage{textcomp}
\usepackage{soul}
\usepackage{xcolor}

\begin{document}
\preprint{APS/123-QED}

\title{Diffuse Inelastic Neutron Scattering from Anharmonic Vibrations in Cuprite}
\author{C. N. Saunders$^{\S}$}
\email{clairenicolesaunders@gmail.com}
\author{V. V. Ladygin$^\S$}
\affiliation{California Institute of Technology, Pasadena, California 91125, USA}
\author{D. S. Kim}
\affiliation{The University of California, Los Angeles, Los Angeles, California 90095, USA}
\author{C. M. Bernal-Choban}
\author{S. H. Lohaus}
\affiliation{California Institute of Technology, Pasadena, California 91125, USA}
\author{G. E. Granroth}
\author{D. L. Abernathy}
\affiliation{Oak Ridge National Laboratory, Oak Ridge, Tennessee 37831, USA}
\author{B. Fultz}
\email{btf@caltech.edu}
\affiliation{California Institute of Technology, Pasadena, California 91125, USA}

\date{\today}

\begin{abstract}
Atomic vibrational dynamics in cuprite, Cu$_2$O, was studied by inelastic neutron scattering and molecular dynamics (MD) simulations from 10\,K to 900\,K. 
Above 300\,K, a diffuse inelastic intensity (DII) appeared, obscuring the high-energy phonon modes. 
Classical MD simulations with a machine learning interatomic potential reproduced general features of the DII, especially with a Langevin thermostat. The DII originates from random phase shifts of vibrating O-atoms that have anharmonic interactions with neighboring Cu-atoms.
\end{abstract}

\maketitle
\def\thefootnote{$\S$}\footnotetext{Equally contributing authors.}

    A significant advance in measuring structural features of materials  occurred with methods of total elastic scattering, which cover a wide range of wavevectors $\vec{Q}$ and include the diffuse elastic intensity between and through the Bragg diffractions \cite{Egami2012}. The Fourier transform of the total elastic scattering gives the real-space pair correlation function of atom positions, revealing short-range atomic structures not available from  sharp Bragg diffraction peaks. 

Historically, inelastic neutron scattering (INS) measurements of crystals have focused on sharp features in the energy spectra $S(\vec{Q}, \omega)$, such as the central frequencies of phonons, $\omega$, and their dispersions, $\omega(\vec{Q})$. A harmonic analysis approximates the phonon dynamics as normal modes of vibration without interactions and therefore long lifetimes. An anharmonic analysis that includes phonon-phonon interactions through cubic perturbations gives broadened spectral features from shortened phonon lifetimes. This spectral broadening in $\omega$ is analogous to the broadening of diffraction peaks in $\vec{Q}$ caused by short crystal dimensions \cite{Olds2018}. 

The phase of an atom vibration is altered if it interacts with neighboring atoms through non-harmonic forces that change with interatomic separations. The timing or magnitude of these phase shifts could have randomness. For comparison, phase errors for elastic scattering from atoms with displacement disorder cause diffuse elastic intensity over a wide range of $Q$.
By analogy, randomness in  phase shifts should generate a diffuse inelastic intensity in $\omega$. 

Here we report diffuse inelastic intensity (DII)  in INS measurements of phonon spectra of cuprite.
In a recent study on phonons and anomalous thermal expansion of cuprite, Cu$_2$O \cite{Saunders2022}, experimental data showed a substantial broadening of the energies of optical modes at 300\,K. This broadening was beyond the anharmonic effects from three-phonon processes. 
At the temperatures of 700\,K and 900\,K reported here, the optical modes are no longer visible as distinct peaks, and the spectral intensity at high energies appears as DII. 
Molecular dynamics (MD) simulations on cuprite under comparable conditions reproduce the DII.
Our interpretation is that of a vibrational mode coupled to a noise source, modeled with the  Schr\"{o}dinger--Langevin equation. 

The INS measurements were performed with the time-of-flight wide angular-range chopper spectrometer, ARCS \cite{Abernathy2012}, at the Spallation Neutron Source \cite{Mason2006}. Phonon dispersions at 10\,K and 300\,K from the same single crystal were presented previously in Ref. \cite{Saunders2022}. The data from 700\,K and 900\,K 
are presented here for the first time. Further details on the experiment and data post-processing are in \cite{Saunders2022} and in the {Supplemental} [URL will be inserted by the production group].

\begin{figure*}
\centering
\includegraphics[width=\textwidth]{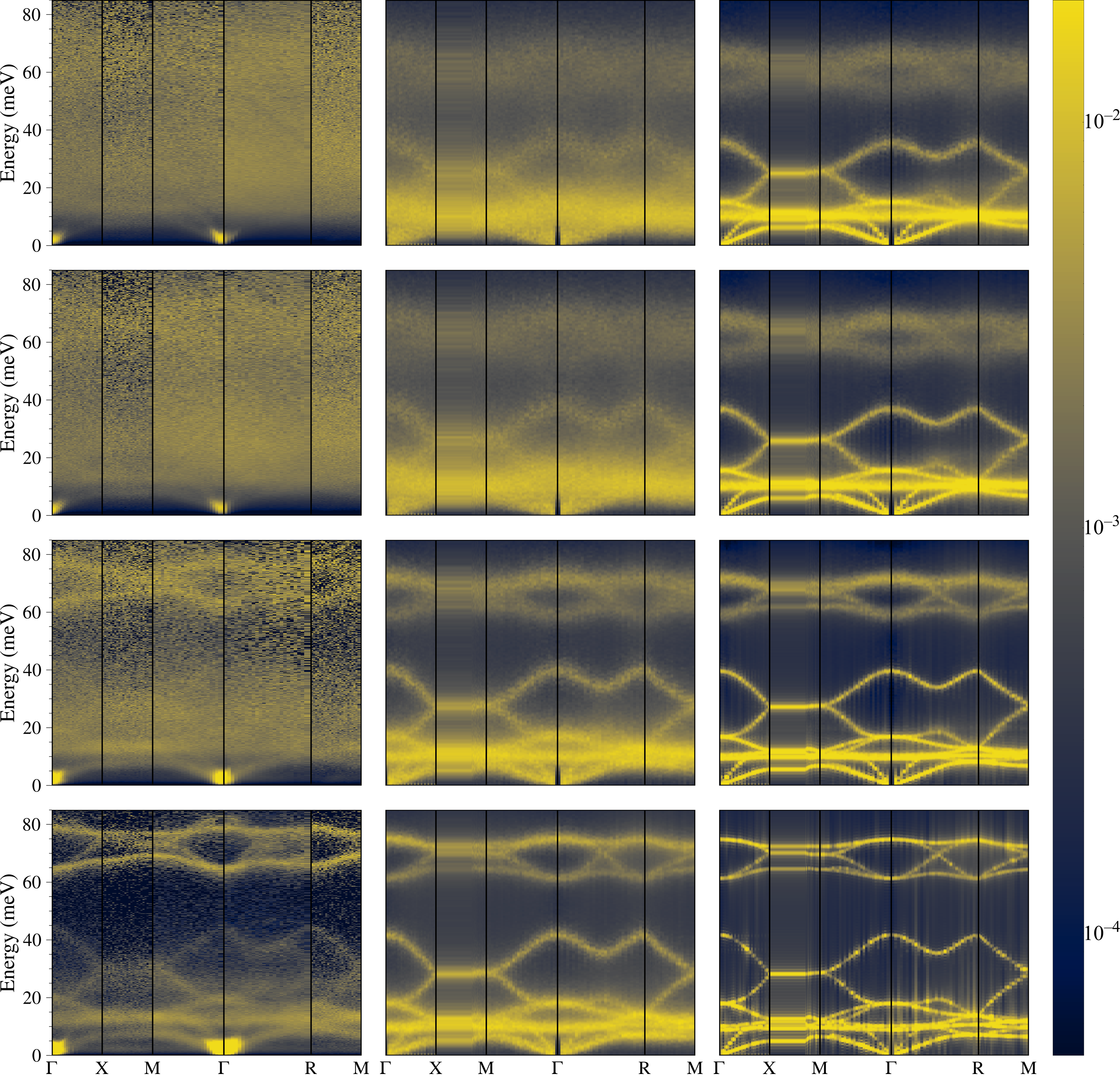}
\caption{Columns from left to right: INS data, MD MLIP with a Langevin thermostat set with $\gamma / \tau_O$= 2 (900\,K and 700\,K), and $\gamma / \tau_O$= 6 (300\,K and 10\,K); and MD \textsc{MLIP} without a Langevin thermostat. Rows from top to bottom: 900\,K, 700\,K, 300\,K, 10\,K. The INS data at 10\,K and 300\,K appears in Ref. \cite{Saunders2022}. Each panel is normalized with respect to its total inelastic intensity (data are on $\log$ scale at right).}
\label{fig:AllExperDisps}
\end{figure*}

The first column of Fig. \ref{fig:AllExperDisps} shows experimental phonon dispersions along some high-symmetry crystallographic directions after folding to include all usable Brillouin zones in the data sets. The lower-energy acoustic dispersions are qualitatively similar at all temperatures. At higher temperatures, the dispersions show obvious changes  at energies above {20}\,meV. Even at 300\,K, there is a significant broadening of the highest optical branches, but at 900\,K only a diffuse, featureless band of inelastic intensity is seen at energies above 40\,meV. This DII is not a deficiency of data processing; the acoustic dispersions were analyzed simultaneously and retained much of their structure at high temperatures. (The high temperatures did not alter the sample, either. The 10\,K data were acquired last.) 

The anharmonic vibrational dynamics were first calculated with the method used in our previous work at lower temperatures \cite{Saunders2022}. Here a stochastic temperature-dependent effective potential (sTDEP) \cite{Hellman2011, Hellman2013_1, Hellman2013_2, Kim2018} was used to fit an ensemble average of the Hellman-Feynman forces in an \textit{ab initio} simulation of supercells with thermally-displaced atoms. The model potential accounts explicitly for cubic anharmonicity and includes the effects of quartic anharmonicity through a renormalization of the harmonic potential. The {Supplemental} and Ref. \cite{Kim2018} explain how the real and imaginary parts of the cubic self-energy correction modify the phonon spectra, following many-body perturbation theory \cite{Maradudin1962, Wallace1998}. Results in the {Supplemental} show that sTDEP does not account for all the DII seen in the experimental data of Fig. \ref{fig:AllExperDisps}. 

The DII is most prominent at high temperatures, so classical molecular dynamics (MD) was used for a second, independent, computational effort. We obtained the spectral intensity through the Fourier transform of a  velocity-velocity autocorrelation function, projected onto $\vec{Q}$. These MD calculations required long lengths and long times for good resolution of phonon features in $\vec{Q}$ and $\omega$ \cite{Lahnsteiner2022}. These simulations were performed with a machine learning interatomic potential (MLIP), trained with input from density functional theory (DFT) calculations performed on supercells with VASP \cite{Kresse1993, Kresse1994, Kresse1996_1, Kresse1996_2}. The DFT calculations used plane-wave basis sets, projector augmented wave (PAW) pseudopotentials, exchange-correlation functionals in the generalized gradient approximation (GGA) \cite{Kresse1999, Perdew1996}, and supercells  generated with thermally-displaced atoms. These training supercells were $3\times3\times3$, containing 162 atoms. The $k$-grid was $2\times2\times2$, and the kinetic energy cutoff was 520\,eV. There were 523 supercells in the training set. 

Machine-learned moment-tensor potentials \cite{Shapeev2016} were used to model the interatomic interaction in classical MD simulations with the LAMMPS code \cite{LAMMPS}. The supercell size was $30\times30\times30$ with 162,000 atoms. The projected velocity-velocity autocorrelation functions came from 20 statistically-independent cuts of {80}\,ps MD trajectories to ensure convergence in the time domain. The results from MD calculations were post-processed in the package VVCORE \cite{Saunders2022_thesis}. The {Supplemental} provides additional details on MLIP construction and computational data analyses. 

The right column of Fig. \ref{fig:AllExperDisps} shows results from microcanonical ensembles after thermal equilibration. The middle column of Fig. \ref{fig:AllExperDisps} shows results from MD simulations with a Langevin thermostat \cite{Schneider1978} that adds a random acceleration to the particle trajectory and damps the particle velocities. The damping parameter was significant ($\gamma/ \tau_{\rm O} = 2$ for 900\,K and 700\,K, where $\tau_{\rm O}$ is the characteristic period of oxygen vibrations in cuprite). The Langevin thermostat is a source of white noise \cite{Pottier2009}, and the large damping parameter corresponds to a strong coupling between vibrational modes and the noise source. Time correlations are shown in the Supplemental. Overall, these full MD simulations account for the experimental spectra at 700\,K and 900\,K with more success than the sTDEP results that use lower-order anharmonicity. 

The broad energy spread of the measured intensity at high temperatures indicates a loss of temporal coherence of atom vibrations and suggests an alternative approach not based on delocalized phonons. 
In the incoherent approximation, the  inelastic intensity is \cite{VanHove1954, Lovesey1984, Squires2012}
\begin{align}
I(\omega) = {\mathcal K} \int_{-\infty}^{\infty} {\mathrm{exp}}({\rm i} \omega t) \langle  \psi^* (t)    \psi(0)  \rangle_{\rm th} \, {\rm d}t \; , \label{ScatteredIntensity}
\end{align}
where $\langle \, \rangle_{\rm th}$  denotes the thermodynamic average and ${\mathcal K} $ is 
\begin{align}
{\mathcal K} =  \frac{k_f}{k_i}\frac{N}{M}\mathrm{exp}({-2W})\frac{{Q}^2}{3}\frac{\sigma}{2\pi\hbar} \; ,
\end{align}
where  \textit{N} is the number of atoms, \textit{M} is the atomic mass, $\sigma$ is the neutron scattering cross section, and $\vec{Q}$ is the momentum transfer. The ratio of final to initial wavevectors, $k_f / k_i$, accounts for the change in flux from the change in the neutron velocity after scattering. The $ \psi(t)$ describes the dynamics of the center-of-mass of a nucleus that scatters the neutron. 

Now we obtain Eq. \ref{ScatteredIntensity} for the inelastic scattering with
the Schr\"{o}dinger--Langevin equation, which is \cite{vanKampen2007}
\begin{align}
\frac{{\rm d} \psi}{{\rm d} t} = -{\rm i} ({\mathcal H} / \hbar) \psi - \frac{1}{2}V^2 \psi + l(t) V \psi \; ,
\end{align}
where ${\mathcal H}$  is the Hamiltonian without fluctuations or damping. Here $V$ is a constant giving the strength of coupling to the fluctuations, which have the time dependence $l(t)$. Having $V$ in the fluctuations (the positive third term on the right) and $V^2$ in
the dissipative damping (negative second term on the right) conserves the norm of $\psi$ \cite{vanKampen2007}. The fluctuation function $l(t)$ has a zero mean, and an autocorrelation function, $P_l(t) $
\begin{align}
\int_{0}^{\infty} l( t) \, {\rm d} t &= 0 \; , \label{AverageZeroNoise} \\
P_l(t) &= \int_{0}^{\infty}  \langle l(t') l(t+t') \rangle_{\rm th}  \, {\rm d} t' \; . \label{Pl(t)}
\end{align}

The inelastic spectrum of Eq. \ref{ScatteredIntensity} requires a product of 
\begin{align}
\psi (0) &=   \psi_0 + l(0) V \psi   \; , \\
\psi^* (t) \! \!  &= \! \!  \psi^* (0) \bigg[ \! \exp  \! \! \left( \! {- {\rm i} \omega_0 t - \frac{1}{2} V^2 t } \right) \! 
+ V \! \!  \!  \int_{0}^{\infty} \! \! \! l(t+t') \, {\rm d} t'  \bigg],
\end{align}
where the energy of the oscillator is $\hbar \omega_0$ and the thermodynamic average is 
\begin{align}
\langle  \psi^*(t) \psi(0) \rangle_{\rm th} &= 
 \mathrm{exp}\left({-{\rm i} \omega_0 t}\right) \mathrm{exp}\left({-{\frac{1}{2} V^2 t } }\right)
 + \ V^2 P_l(t) \; . \label{Agvepsi*psi}
\end{align}
There are no cross terms in Eq. \ref{Agvepsi*psi} owing to Eq. \ref{AverageZeroNoise}, and $\langle \psi^*(0) \psi(0) \rangle = 1$. 

Equation \ref{ScatteredIntensity} for the inelastic intensity then has two terms 
\begin{align}
I(\omega)/{\mathcal K}=  \frac{V^2}{(\omega-\omega_0)^2 + V^4} + V^2 \, \widetilde{\mathcal P}_{{\Delta \omega}}(\omega) \; .
\label{WhiteNoiseIntensity22}
\end{align}
The first term is the familiar spectrum of a damped harmonic oscillator, a Lorentzian function about $\omega_0$ of breadth $V^2$. The second term from the fluctuations scales with the coupling coefficient $V^2$. A simple example is an instantaneous time correlation $P_l(t) = \delta(t)$, for which $\widetilde{\mathcal P}_{{\Delta \omega}}(\omega) =1$, giving a flat, diffuse, white-noise background in the inelastic intensity. This term is new in the context of INS, although a diffuse elastic scattering of disordered materials is analogous \cite{Cowley1995}. 

We expect $P_l(t)$ to peak at $t=0$ with some spread in time, as in the generalized Langevin equation. Its Fourier transform, $\widetilde{\mathcal P}_{{\Delta \omega}}(\omega)$, modulates the DII, giving it a characteristic width ${\Delta \omega}$, and suppressing the DII at large $\omega$. An extended time correlation for $P_l(t)$ gives a smaller characteristic  ${\Delta \omega}$ and more suppression of high frequencies in the inelastic spectrum. 


Our prior work \cite{Saunders2022} showed that the 
partial DOS for O-atoms is a peak at approximately {72}\,meV, consistent
with the rather flat optical branches of cuprite. 
A convenient physical picture is that an O-atom behaves as a local
Einstein oscillator in a cage of four Cu atoms. 
With small vibrational displacements at low temperatures, the interatomic forces 
are predominantly harmonic, and crisp phonon dispersions are seen at 10\,K in Fig. \ref{fig:AllExperDisps}. 
When the atoms vibrate with larger amplitudes, the interatomic force departs from harmonic behavior as the O- and Cu-atom neighbors approach more closely. 
A harmonic oscillation has a linear increase in phase with time.
As the potential energy rises above the harmonic potential during close approaches of the O- and Cu-atoms, the phase is advanced more rapidly than for a harmonic oscillation, especially for local modes of the light O-atoms. 
This phase advance occurs with some randomness in time as the tetrahedral cage of Cu-atoms undergoes deformations. Over time, the vibrational modes for O-atoms recover to an average energy of  $k_{\mathrm{B}}T$, but the phase change is cumulative. 

A random component in the timing of anharmonic interactions generates DII, and a wider distribution of the times of individual phase shifts gives a broader DII. A plausible shape for the low-energy part of the DII spectrum  is shown in Fig. 2 of the {Supplemental}, where semi-periodic interactions are assumed.
The characteristic width ${\Delta \omega}$ of the diffuse intensity in Fig. \ref{fig:AllExperDisps} increases with temperature from 300\,K to 900\,K. This width corresponds to a shorter time correlation, decreasing from approximately  3$\times10^{-13}$\,s ($\gamma / \tau_O$= 6) at 300\,K to 1$\times10^{-13}$\,s ($\gamma / \tau_O$= 2) at 900\,K. For a characteristic frequency of 17\,THz for O-atoms in the upper optical modes, the corresponding spread in phase shifts increases from approximately $2\pi$/10 to $2\pi$/4 per cycle. 
Some time correlations from experiment and computation are shown in the Supplemental. 

The DII in cuprite is fortuitously large and overwhelms the optical dispersions. 
Other anharmonic crystals retain discernible phonon dispersions at high temperatures without an obvious DII. Nevertheless, DII may occur beneath phonon peaks in the spectra of other anharmonic crystals. DII looks like a background from other sources, so previous experimental studies may have discarded it. 
There is evidence of DII in the phonon DOS of Ag$_2$O, which is structurally similar to cuprite and has weakly constrained M-O-M (M=Cu, Ag) bonds \cite{Lan2015}. 
Other crystals that may show such DII include 
skutterudites, clathrate inclusion compounds, and methylammonium lead iodide perovskite.  
Their rattling motions of atoms or molecules within the confines of hard walls may give random departures from harmonic potentials and incoherent phase shifts \cite{Dong2000, Sergueev2015, Ferreira2020}. 
Liquids may also show a form of DII. 

Individual processes of phonon-phonon interactions give anharmonic behaviors such as finite phonon lifetimes. A many-body analysis of three- and four-phonon processes, which has proved effective for calculating anharmonic self-energy corrections, assumes coherent interactions between individual phonons \cite{Maradudin1962, Wallace1998, BarronKlein1974}. The DII arises from random incoherent phase shifts, and may be
more consistent with some concepts of a thermal bath that contributes noise in quantum systems \cite{Gardiner2000, Clerk2010}. 

\begin{acknowledgments}
This research used resources at the Spallation Neutron Source, a DOE Office of Science User Facility operated by the Oak Ridge National Laboratory. In addition, this work used resources from the National Energy Research Scientific Computing Center (NERSC), a DOE Office of Science User Facility supported by the Office of Science of the US Department of Energy under Contract DE-AC02-05CH11231. Finally, this work was supported by the DOE Office of Science, BES, under Contract DE-FG02-03ER46055.
\end{acknowledgments}

\bibliography{main.bib}

\begin{thebibliography}{35}%
\makeatletter
\providecommand \@ifxundefined [1]{%
 \@ifx{#1\undefined}
}%
\providecommand \@ifnum [1]{%
 \ifnum #1\expandafter \@firstoftwo
 \else \expandafter \@secondoftwo
 \fi
}%
\providecommand \@ifx [1]{%
 \ifx #1\expandafter \@firstoftwo
 \else \expandafter \@secondoftwo
 \fi
}%
\providecommand \natexlab [1]{#1}%
\providecommand \enquote  [1]{``#1''}%
\providecommand \bibnamefont  [1]{#1}%
\providecommand \bibfnamefont [1]{#1}%
\providecommand \citenamefont [1]{#1}%
\providecommand \href@noop [0]{\@secondoftwo}%
\providecommand \href [0]{\begingroup \@sanitize@url \@href}%
\providecommand \@href[1]{\@@startlink{#1}\@@href}%
\providecommand \@@href[1]{\endgroup#1\@@endlink}%
\providecommand \@sanitize@url [0]{\catcode `\\12\catcode `\$12\catcode
  `\&12\catcode `\#12\catcode `\^12\catcode `\_12\catcode `\%12\relax}%
\providecommand \@@startlink[1]{}%
\providecommand \@@endlink[0]{}%
\providecommand \url  [0]{\begingroup\@sanitize@url \@url }%
\providecommand \@url [1]{\endgroup\@href {#1}{\urlprefix }}%
\providecommand \urlprefix  [0]{URL }%
\providecommand \Eprint [0]{\href }%
\providecommand \doibase [0]{https://doi.org/}%
\providecommand \selectlanguage [0]{\@gobble}%
\providecommand \bibinfo  [0]{\@secondoftwo}%
\providecommand \bibfield  [0]{\@secondoftwo}%
\providecommand \translation [1]{[#1]}%
\providecommand \BibitemOpen [0]{}%
\providecommand \bibitemStop [0]{}%
\providecommand \bibitemNoStop [0]{.\EOS\space}%
\providecommand \EOS [0]{\spacefactor3000\relax}%
\providecommand \BibitemShut  [1]{\csname bibitem#1\endcsname}%
\let\auto@bib@innerbib\@empty
\bibitem [{\citenamefont {Egami}\ and\ \citenamefont
  {Billinge}(2012)}]{Egami2012}%
  \BibitemOpen
  \bibfield  {author} {\bibinfo {author} {\bibfnamefont {T.}~\bibnamefont
  {Egami}}\ and\ \bibinfo {author} {\bibfnamefont {S.~J.~L.}\ \bibnamefont
  {Billinge}},\ }\href
  {https://doi.org/https://doi.org/10.1016/B978-0-08-097133-9.09992-5} {\emph
  {\bibinfo {title} {Underneath the {B}ragg {P}eaks}}},\ \bibinfo {series}
  {Pergamon Materials Series}, Vol.~\bibinfo {volume} {16}\ (\bibinfo
  {publisher} {Pergamon},\ \bibinfo {year} {2012})\BibitemShut {NoStop}%
\bibitem [{\citenamefont {Olds}\ \emph {et~al.}(2018)\citenamefont {Olds},
  \citenamefont {Saunders}, \citenamefont {Peters}, \citenamefont {Proffen},
  \citenamefont {Neuefeind},\ and\ \citenamefont {Page}}]{Olds2018}%
  \BibitemOpen
  \bibfield  {author} {\bibinfo {author} {\bibfnamefont {D.}~\bibnamefont
  {Olds}}, \bibinfo {author} {\bibfnamefont {C.~N.}\ \bibnamefont {Saunders}},
  \bibinfo {author} {\bibfnamefont {M.}~\bibnamefont {Peters}}, \bibinfo
  {author} {\bibfnamefont {T.}~\bibnamefont {Proffen}}, \bibinfo {author}
  {\bibfnamefont {J.}~\bibnamefont {Neuefeind}},\ and\ \bibinfo {author}
  {\bibfnamefont {K.}~\bibnamefont {Page}},\ }\href
  {https://doi.org/10.1107/S2053273318003224} {\bibfield  {journal} {\bibinfo
  {journal} {Acta. Crystallogr. A}\ }\textbf {\bibinfo {volume} {74}},\
  \bibinfo {pages} {293} (\bibinfo {year} {2018})}\BibitemShut {NoStop}%
\bibitem [{\citenamefont {Saunders}\ \emph {et~al.}(2022)\citenamefont
  {Saunders}, \citenamefont {Kim}, \citenamefont {Hellman}, \citenamefont
  {Smith}, \citenamefont {Weadock}, \citenamefont {Omelchenko}, \citenamefont
  {Granroth}, \citenamefont {Bernal-Choban}, \citenamefont {Lohaus},
  \citenamefont {Abernathy},\ and\ \citenamefont {Fultz}}]{Saunders2022}%
  \BibitemOpen
  \bibfield  {author} {\bibinfo {author} {\bibfnamefont {C.~N.}\ \bibnamefont
  {Saunders}}, \bibinfo {author} {\bibfnamefont {D.~S.}\ \bibnamefont {Kim}},
  \bibinfo {author} {\bibfnamefont {O.}~\bibnamefont {Hellman}}, \bibinfo
  {author} {\bibfnamefont {H.~L.}\ \bibnamefont {Smith}}, \bibinfo {author}
  {\bibfnamefont {N.~J.}\ \bibnamefont {Weadock}}, \bibinfo {author}
  {\bibfnamefont {S.~T.}\ \bibnamefont {Omelchenko}}, \bibinfo {author}
  {\bibfnamefont {G.~E.}\ \bibnamefont {Granroth}}, \bibinfo {author}
  {\bibfnamefont {C.~M.}\ \bibnamefont {Bernal-Choban}}, \bibinfo {author}
  {\bibfnamefont {S.~H.}\ \bibnamefont {Lohaus}}, \bibinfo {author}
  {\bibfnamefont {D.~L.}\ \bibnamefont {Abernathy}},\ and\ \bibinfo {author}
  {\bibfnamefont {B.}~\bibnamefont {Fultz}},\ }\href
  {https://doi.org/10.1103/PhysRevB.105.174308} {\bibfield  {journal} {\bibinfo
   {journal} {Phys. Rev. B}\ }\textbf {\bibinfo {volume} {105}},\ \bibinfo
  {pages} {174308} (\bibinfo {year} {2022})}\BibitemShut {NoStop}%
\bibitem [{\citenamefont {Abernathy}\ \emph {et~al.}(2012)\citenamefont
  {Abernathy}, \citenamefont {Stone}, \citenamefont {Loguillo}, \citenamefont
  {Lucas}, \citenamefont {Delaire}, \citenamefont {Tang}, \citenamefont {Lin},\
  and\ \citenamefont {Fultz}}]{Abernathy2012}%
  \BibitemOpen
  \bibfield  {author} {\bibinfo {author} {\bibfnamefont {D.~L.}\ \bibnamefont
  {Abernathy}}, \bibinfo {author} {\bibfnamefont {M.~B.}\ \bibnamefont
  {Stone}}, \bibinfo {author} {\bibfnamefont {M.~J.}\ \bibnamefont {Loguillo}},
  \bibinfo {author} {\bibfnamefont {M.~S.}\ \bibnamefont {Lucas}}, \bibinfo
  {author} {\bibfnamefont {O.}~\bibnamefont {Delaire}}, \bibinfo {author}
  {\bibfnamefont {X.}~\bibnamefont {Tang}}, \bibinfo {author} {\bibfnamefont
  {J.~Y.~Y.}\ \bibnamefont {Lin}},\ and\ \bibinfo {author} {\bibfnamefont
  {B.}~\bibnamefont {Fultz}},\ }\href {https://doi.org/10.1063/1.3680104}
  {\bibfield  {journal} {\bibinfo  {journal} {Rev. Sci. Instrum.}\ }\textbf
  {\bibinfo {volume} {83}},\ \bibinfo {pages} {015114} (\bibinfo {year}
  {2012})}\BibitemShut {NoStop}%
\bibitem [{\citenamefont {Mason}\ \emph {et~al.}(2006)\citenamefont {Mason},
  \citenamefont {Abernathy}, \citenamefont {Anderson}, \citenamefont {Ankner},
  \citenamefont {Egami}, \citenamefont {Ehlers}, \citenamefont {Ekkebus},
  \citenamefont {Granroth}, \citenamefont {Hagen}, \citenamefont {Herwig},
  \citenamefont {Hodges}, \citenamefont {Hoffmann}, \citenamefont {Horak},
  \citenamefont {Horton}, \citenamefont {Klose}, \citenamefont {Larese},
  \citenamefont {Mesecar}, \citenamefont {Myles}, \citenamefont {Neuefeind},
  \citenamefont {Ohl}, \citenamefont {Tulk}, \citenamefont {Wang},\ and\
  \citenamefont {Zhao}}]{Mason2006}%
  \BibitemOpen
  \bibfield  {author} {\bibinfo {author} {\bibfnamefont {T.}~\bibnamefont
  {Mason}}, \bibinfo {author} {\bibfnamefont {D.}~\bibnamefont {Abernathy}},
  \bibinfo {author} {\bibfnamefont {I.}~\bibnamefont {Anderson}}, \bibinfo
  {author} {\bibfnamefont {J.}~\bibnamefont {Ankner}}, \bibinfo {author}
  {\bibfnamefont {T.}~\bibnamefont {Egami}}, \bibinfo {author} {\bibfnamefont
  {G.}~\bibnamefont {Ehlers}}, \bibinfo {author} {\bibfnamefont
  {A.}~\bibnamefont {Ekkebus}}, \bibinfo {author} {\bibfnamefont
  {G.}~\bibnamefont {Granroth}}, \bibinfo {author} {\bibfnamefont
  {M.}~\bibnamefont {Hagen}}, \bibinfo {author} {\bibfnamefont
  {K.}~\bibnamefont {Herwig}}, \bibinfo {author} {\bibfnamefont
  {J.}~\bibnamefont {Hodges}}, \bibinfo {author} {\bibfnamefont
  {C.}~\bibnamefont {Hoffmann}}, \bibinfo {author} {\bibfnamefont
  {C.}~\bibnamefont {Horak}}, \bibinfo {author} {\bibfnamefont
  {L.}~\bibnamefont {Horton}}, \bibinfo {author} {\bibfnamefont
  {F.}~\bibnamefont {Klose}}, \bibinfo {author} {\bibfnamefont
  {J.}~\bibnamefont {Larese}}, \bibinfo {author} {\bibfnamefont
  {A.}~\bibnamefont {Mesecar}}, \bibinfo {author} {\bibfnamefont
  {D.}~\bibnamefont {Myles}}, \bibinfo {author} {\bibfnamefont
  {J.}~\bibnamefont {Neuefeind}}, \bibinfo {author} {\bibfnamefont
  {M.}~\bibnamefont {Ohl}}, \bibinfo {author} {\bibfnamefont {C.}~\bibnamefont
  {Tulk}}, \bibinfo {author} {\bibfnamefont {X.-L.}\ \bibnamefont {Wang}},\
  and\ \bibinfo {author} {\bibfnamefont {J.}~\bibnamefont {Zhao}},\ }\href
  {https://doi.org/10.1016/j.physb.2006.05.281} {\bibfield  {journal} {\bibinfo
   {journal} {Physica B Condens. Matter}\ }\textbf {\bibinfo {volume}
  {385-386}},\ \bibinfo {pages} {955} (\bibinfo {year} {2006})}\BibitemShut
  {NoStop}%
\bibitem [{\citenamefont {Hellman}\ \emph {et~al.}(2011)\citenamefont
  {Hellman}, \citenamefont {Abrikosov},\ and\ \citenamefont
  {Simak}}]{Hellman2011}%
  \BibitemOpen
  \bibfield  {author} {\bibinfo {author} {\bibfnamefont {O.}~\bibnamefont
  {Hellman}}, \bibinfo {author} {\bibfnamefont {I.~A.}\ \bibnamefont
  {Abrikosov}},\ and\ \bibinfo {author} {\bibfnamefont {S.~I.}\ \bibnamefont
  {Simak}},\ }\href {https://doi.org/10.1103/PhysRevB.84.180301} {\bibfield
  {journal} {\bibinfo  {journal} {Phys. Rev. B}\ }\textbf {\bibinfo {volume}
  {84}},\ \bibinfo {pages} {180301} (\bibinfo {year} {2011})}\BibitemShut
  {NoStop}%
\bibitem [{\citenamefont {Hellman}\ \emph {et~al.}(2013)\citenamefont
  {Hellman}, \citenamefont {Stenetegand}, \citenamefont {Abrikosov},\ and\
  \citenamefont {Simak}}]{Hellman2013_1}%
  \BibitemOpen
  \bibfield  {author} {\bibinfo {author} {\bibfnamefont {O.}~\bibnamefont
  {Hellman}}, \bibinfo {author} {\bibfnamefont {P.}~\bibnamefont
  {Stenetegand}}, \bibinfo {author} {\bibfnamefont {I.~A.}\ \bibnamefont
  {Abrikosov}},\ and\ \bibinfo {author} {\bibfnamefont {S.~I.}\ \bibnamefont
  {Simak}},\ }\href {https://doi.org/10.1103/PhysRevB.87.104111} {\bibfield
  {journal} {\bibinfo  {journal} {Phys. Rev. B}\ }\textbf {\bibinfo {volume}
  {87}},\ \bibinfo {pages} {104111} (\bibinfo {year} {2013})}\BibitemShut
  {NoStop}%
\bibitem [{\citenamefont {Hellman}\ and\ \citenamefont
  {Abrikosov}(2013)}]{Hellman2013_2}%
  \BibitemOpen
  \bibfield  {author} {\bibinfo {author} {\bibfnamefont {O.}~\bibnamefont
  {Hellman}}\ and\ \bibinfo {author} {\bibfnamefont {I.~A.}\ \bibnamefont
  {Abrikosov}},\ }\href {https://doi.org/10.1103/PhysRevB.88.144301} {\bibfield
   {journal} {\bibinfo  {journal} {Phys. Rev. B}\ }\textbf {\bibinfo {volume}
  {88}},\ \bibinfo {pages} {144301} (\bibinfo {year} {2013})}\BibitemShut
  {NoStop}%
\bibitem [{\citenamefont {Kim}\ \emph {et~al.}(2018)\citenamefont {Kim},
  \citenamefont {Hellman}, \citenamefont {Herriman}, \citenamefont {Smith},
  \citenamefont {Lin}, \citenamefont {Shulumba}, \citenamefont {Niedziela},
  \citenamefont {Li}, \citenamefont {Abernathy},\ and\ \citenamefont
  {Fultz}}]{Kim2018}%
  \BibitemOpen
  \bibfield  {author} {\bibinfo {author} {\bibfnamefont {D.~S.}\ \bibnamefont
  {Kim}}, \bibinfo {author} {\bibfnamefont {O.}~\bibnamefont {Hellman}},
  \bibinfo {author} {\bibfnamefont {J.}~\bibnamefont {Herriman}}, \bibinfo
  {author} {\bibfnamefont {H.~L.}\ \bibnamefont {Smith}}, \bibinfo {author}
  {\bibfnamefont {J.~Y.~Y.}\ \bibnamefont {Lin}}, \bibinfo {author}
  {\bibfnamefont {N.}~\bibnamefont {Shulumba}}, \bibinfo {author}
  {\bibfnamefont {J.~L.}\ \bibnamefont {Niedziela}}, \bibinfo {author}
  {\bibfnamefont {C.~W.}\ \bibnamefont {Li}}, \bibinfo {author} {\bibfnamefont
  {D.~L.}\ \bibnamefont {Abernathy}},\ and\ \bibinfo {author} {\bibfnamefont
  {B.}~\bibnamefont {Fultz}},\ }\href {https://doi.org/10.1073/pnas.1707745115}
  {\bibfield  {journal} {\bibinfo  {journal} {Proc. Natl. Acad. Sci. USA}\
  }\textbf {\bibinfo {volume} {115}},\ \bibinfo {pages} {1992} (\bibinfo {year}
  {2018})}\BibitemShut {NoStop}%
\bibitem [{\citenamefont {Maradudin}\ and\ \citenamefont
  {Fein}(1962)}]{Maradudin1962}%
  \BibitemOpen
  \bibfield  {author} {\bibinfo {author} {\bibfnamefont {A.~A.}\ \bibnamefont
  {Maradudin}}\ and\ \bibinfo {author} {\bibfnamefont {A.~E.}\ \bibnamefont
  {Fein}},\ }\href {https://doi.org/10.1103/PhysRev.128.2589} {\bibfield
  {journal} {\bibinfo  {journal} {Phys. Rev.}\ }\textbf {\bibinfo {volume}
  {128}},\ \bibinfo {pages} {2589} (\bibinfo {year} {1962})}\BibitemShut
  {NoStop}%
\bibitem [{\citenamefont {Wallace}(1998)}]{Wallace1998}%
  \BibitemOpen
  \bibfield  {author} {\bibinfo {author} {\bibfnamefont {D.~C.}\ \bibnamefont
  {Wallace}},\ }\href@noop {} {\emph {\bibinfo {title} {Thermodynamics of
  Crystals}}}\ (\bibinfo  {publisher} {J. Wiley, New York},\ \bibinfo {year}
  {1998})\BibitemShut {NoStop}%
\bibitem [{\citenamefont {Lahnsteiner}\ and\ \citenamefont
  {Bokdam}(2022)}]{Lahnsteiner2022}%
  \BibitemOpen
  \bibfield  {author} {\bibinfo {author} {\bibfnamefont {J.}~\bibnamefont
  {Lahnsteiner}}\ and\ \bibinfo {author} {\bibfnamefont {M.}~\bibnamefont
  {Bokdam}},\ }\href {https://doi.org/10.1103/PhysRevB.105.024302} {\bibfield
  {journal} {\bibinfo  {journal} {Phys. Rev. B}\ }\textbf {\bibinfo {volume}
  {105}},\ \bibinfo {pages} {024302} (\bibinfo {year} {2022})}\BibitemShut
  {NoStop}%
\bibitem [{\citenamefont {Kresse}\ and\ \citenamefont
  {Hafner}(1993)}]{Kresse1993}%
  \BibitemOpen
  \bibfield  {author} {\bibinfo {author} {\bibfnamefont {G.}~\bibnamefont
  {Kresse}}\ and\ \bibinfo {author} {\bibfnamefont {J.}~\bibnamefont
  {Hafner}},\ }\href {https://doi.org/10.1103/PhysRevB.47.558} {\bibfield
  {journal} {\bibinfo  {journal} {Phys. Rev. B}\ }\textbf {\bibinfo {volume}
  {47}},\ \bibinfo {pages} {558} (\bibinfo {year} {1993})}\BibitemShut
  {NoStop}%
\bibitem [{\citenamefont {Kresse}\ and\ \citenamefont
  {Hafner}(1994)}]{Kresse1994}%
  \BibitemOpen
  \bibfield  {author} {\bibinfo {author} {\bibfnamefont {G.}~\bibnamefont
  {Kresse}}\ and\ \bibinfo {author} {\bibfnamefont {J.}~\bibnamefont
  {Hafner}},\ }\href {https://doi.org/10.1103/PhysRevB.49.14251} {\bibfield
  {journal} {\bibinfo  {journal} {Phys. Rev. B}\ }\textbf {\bibinfo {volume}
  {49}},\ \bibinfo {pages} {14251} (\bibinfo {year} {1994})}\BibitemShut
  {NoStop}%
\bibitem [{\citenamefont {Kresse}\ and\ \citenamefont
  {Furthm{\"u}ller}(1996)}]{Kresse1996_1}%
  \BibitemOpen
  \bibfield  {author} {\bibinfo {author} {\bibfnamefont {G.}~\bibnamefont
  {Kresse}}\ and\ \bibinfo {author} {\bibfnamefont {J.}~\bibnamefont
  {Furthm{\"u}ller}},\ }\href {https://doi.org/10.1016/0927-0256(96)00008-0}
  {\bibfield  {journal} {\bibinfo  {journal} {Comput. Mater. Sci.}\ }\textbf
  {\bibinfo {volume} {6}},\ \bibinfo {pages} {15 } (\bibinfo {year}
  {1996})}\BibitemShut {NoStop}%
\bibitem [{\citenamefont {Kresse}\ and\ \citenamefont
  {Furthm\"uller}(1996)}]{Kresse1996_2}%
  \BibitemOpen
  \bibfield  {author} {\bibinfo {author} {\bibfnamefont {G.}~\bibnamefont
  {Kresse}}\ and\ \bibinfo {author} {\bibfnamefont {J.}~\bibnamefont
  {Furthm\"uller}},\ }\href {https://doi.org/10.1103/PhysRevB.54.11169}
  {\bibfield  {journal} {\bibinfo  {journal} {Phys. Rev. B}\ }\textbf {\bibinfo
  {volume} {54}},\ \bibinfo {pages} {11169} (\bibinfo {year}
  {1996})}\BibitemShut {NoStop}%
\bibitem [{\citenamefont {Kresse}\ and\ \citenamefont
  {Joubert}(1999)}]{Kresse1999}%
  \BibitemOpen
  \bibfield  {author} {\bibinfo {author} {\bibfnamefont {G.}~\bibnamefont
  {Kresse}}\ and\ \bibinfo {author} {\bibfnamefont {D.}~\bibnamefont
  {Joubert}},\ }\href {https://doi.org/10.1103/PhysRevB.59.1758} {\bibfield
  {journal} {\bibinfo  {journal} {Phys. Rev. B}\ }\textbf {\bibinfo {volume}
  {59}},\ \bibinfo {pages} {1758} (\bibinfo {year} {1999})}\BibitemShut
  {NoStop}%
\bibitem [{\citenamefont {Perdew}\ \emph {et~al.}(1996)\citenamefont {Perdew},
  \citenamefont {Burke},\ and\ \citenamefont {Ernzerhof}}]{Perdew1996}%
  \BibitemOpen
  \bibfield  {author} {\bibinfo {author} {\bibfnamefont {J.~P.}\ \bibnamefont
  {Perdew}}, \bibinfo {author} {\bibfnamefont {K.}~\bibnamefont {Burke}},\ and\
  \bibinfo {author} {\bibfnamefont {M.}~\bibnamefont {Ernzerhof}},\ }\href
  {https://doi.org/10.1103/PhysRevLett.77.3865} {\bibfield  {journal} {\bibinfo
   {journal} {Phys. Rev. Lett.}\ }\textbf {\bibinfo {volume} {77}},\ \bibinfo
  {pages} {3865} (\bibinfo {year} {1996})}\BibitemShut {NoStop}%
\bibitem [{\citenamefont {Shapeev}(2016)}]{Shapeev2016}%
  \BibitemOpen
  \bibfield  {author} {\bibinfo {author} {\bibfnamefont {A.~V.}\ \bibnamefont
  {Shapeev}},\ }\href {https://doi.org/10.1137/15M1054183} {\bibfield
  {journal} {\bibinfo  {journal} {Multiscale Modeling \& Simulation}\ }\textbf
  {\bibinfo {volume} {14}},\ \bibinfo {pages} {1153} (\bibinfo {year}
  {2016})}\BibitemShut {NoStop}%
\bibitem [{\citenamefont {Thompson}\ \emph {et~al.}(2022)\citenamefont
  {Thompson}, \citenamefont {Aktulga}, \citenamefont {Berger}, \citenamefont
  {Bolintineanu}, \citenamefont {Brown}, \citenamefont {Crozier}, \citenamefont
  {{in't Veld}}, \citenamefont {Kohlmeyer}, \citenamefont {Moore},
  \citenamefont {Nguyen}, \citenamefont {Shan}, \citenamefont {Stevens},
  \citenamefont {Tranchida}, \citenamefont {Trott},\ and\ \citenamefont
  {Plimpton}}]{LAMMPS}%
  \BibitemOpen
  \bibfield  {author} {\bibinfo {author} {\bibfnamefont {A.~P.}\ \bibnamefont
  {Thompson}}, \bibinfo {author} {\bibfnamefont {H.~M.}\ \bibnamefont
  {Aktulga}}, \bibinfo {author} {\bibfnamefont {R.}~\bibnamefont {Berger}},
  \bibinfo {author} {\bibfnamefont {D.~S.}\ \bibnamefont {Bolintineanu}},
  \bibinfo {author} {\bibfnamefont {W.~M.}\ \bibnamefont {Brown}}, \bibinfo
  {author} {\bibfnamefont {P.~S.}\ \bibnamefont {Crozier}}, \bibinfo {author}
  {\bibfnamefont {P.~J.}\ \bibnamefont {{in't Veld}}}, \bibinfo {author}
  {\bibfnamefont {A.}~\bibnamefont {Kohlmeyer}}, \bibinfo {author}
  {\bibfnamefont {S.~G.}\ \bibnamefont {Moore}}, \bibinfo {author}
  {\bibfnamefont {T.~D.}\ \bibnamefont {Nguyen}}, \bibinfo {author}
  {\bibfnamefont {R.}~\bibnamefont {Shan}}, \bibinfo {author} {\bibfnamefont
  {M.~J.}\ \bibnamefont {Stevens}}, \bibinfo {author} {\bibfnamefont
  {J.}~\bibnamefont {Tranchida}}, \bibinfo {author} {\bibfnamefont
  {C.}~\bibnamefont {Trott}},\ and\ \bibinfo {author} {\bibfnamefont {S.~J.}\
  \bibnamefont {Plimpton}},\ }\href {https://doi.org/10.1016/j.cpc.2021.108171}
  {\bibfield  {journal} {\bibinfo  {journal} {Comp. Phys. Comm.}\ }\textbf
  {\bibinfo {volume} {271}},\ \bibinfo {pages} {108171} (\bibinfo {year}
  {2022})}\BibitemShut {NoStop}%
\bibitem [{\citenamefont {Saunders}(2022)}]{Saunders2022_thesis}%
  \BibitemOpen
  \bibfield  {author} {\bibinfo {author} {\bibfnamefont {C.~N.}\ \bibnamefont
  {Saunders}},\ }\emph {\bibinfo {title} {Thermal Behavior of Cuprous Oxide: a
  Comprehensive Study of Three-Body Phonon Effects and Beyond}},\ \href
  {https://doi.org/https://doi.org/10.7907/mate-2v65} {Ph.D. thesis},\ \bibinfo
   {school} {California Institute of Technology} (\bibinfo {year}
  {2022})\BibitemShut {NoStop}%
\bibitem [{\citenamefont {Schneider}\ and\ \citenamefont
  {Stoll}(1978)}]{Schneider1978}%
  \BibitemOpen
  \bibfield  {author} {\bibinfo {author} {\bibfnamefont {T.}~\bibnamefont
  {Schneider}}\ and\ \bibinfo {author} {\bibfnamefont {E.}~\bibnamefont
  {Stoll}},\ }\href {https://doi.org/10.1103/PhysRevB.17.1302} {\bibfield
  {journal} {\bibinfo  {journal} {Phys. Rev. B}\ }\textbf {\bibinfo {volume}
  {17}},\ \bibinfo {pages} {1302} (\bibinfo {year} {1978})}\BibitemShut
  {NoStop}%
\bibitem [{\citenamefont {Pottier}(2009)}]{Pottier2009}%
  \BibitemOpen
  \bibfield  {author} {\bibinfo {author} {\bibfnamefont {N.}~\bibnamefont
  {Pottier}},\ }\href@noop {} {\emph {\bibinfo {title} {Nonequilibrium
  {S}tatistical {P}hysics: {L}inear {I}rreversible {P}rocesses}}}\ (\bibinfo
  {publisher} {Oxford University Press},\ \bibinfo {year} {2009})\BibitemShut
  {NoStop}%
\bibitem [{\citenamefont {{Van Hove}}(1954)}]{VanHove1954}%
  \BibitemOpen
  \bibfield  {author} {\bibinfo {author} {\bibfnamefont {L.}~\bibnamefont {{Van
  Hove}}},\ }\href {https://doi.org/10.1103/PhysRev.95.249} {\bibfield
  {journal} {\bibinfo  {journal} {Phys. Rev.}\ }\textbf {\bibinfo {volume}
  {95}},\ \bibinfo {pages} {249} (\bibinfo {year} {1954})}\BibitemShut
  {NoStop}%
\bibitem [{\citenamefont {Lovesey}(1984)}]{Lovesey1984}%
  \BibitemOpen
  \bibfield  {author} {\bibinfo {author} {\bibfnamefont {S.~W.}\ \bibnamefont
  {Lovesey}},\ }\href
  {http://inis.iaea.org/search/search.aspx?orig_q=RN:16036521} {\emph {\bibinfo
  {title} {Theory of {N}eutron {S}cattering from {Co}ndensed {M}atter}}}\
  (\bibinfo  {publisher} {Clarendon Press},\ \bibinfo {address} {United
  Kingdom},\ \bibinfo {year} {1984})\BibitemShut {NoStop}%
\bibitem [{\citenamefont {Squires}(2012)}]{Squires2012}%
  \BibitemOpen
  \bibfield  {author} {\bibinfo {author} {\bibfnamefont {G.~L.}\ \bibnamefont
  {Squires}},\ }\href {https://doi.org/10.1017/CBO9781139107808} {\emph
  {\bibinfo {title} {Introduction to the {T}heory of {T}hermal {N}eutron
  {S}cattering}}},\ \bibinfo {edition} {3rd}\ ed.\ (\bibinfo  {publisher}
  {Cambridge University Press},\ \bibinfo {year} {2012})\BibitemShut {NoStop}%
\bibitem [{\citenamefont {{Van Kampen}}(2007)}]{vanKampen2007}%
  \BibitemOpen
  \bibfield  {author} {\bibinfo {author} {\bibfnamefont {N.~G.}\ \bibnamefont
  {{Van Kampen}}},\ }\href
  {https://doi.org/https://doi.org/10.1016/B978-044452965-7/50020-9} {\emph
  {\bibinfo {title} {Stochastic {P}rocesses in {P}hysics and {C}hemistry}}},\
  \bibinfo {edition} {3rd}\ ed.,\ North-Holland Personal Library\ (\bibinfo
  {publisher} {Elsevier},\ \bibinfo {address} {Amsterdam},\ \bibinfo {year}
  {2007})\ Chap.~\bibinfo {chapter} {17}, pp.\ \bibinfo {pages}
  {422--456}\BibitemShut {NoStop}%
\bibitem [{\citenamefont {Cowley}(1995)}]{Cowley1995}%
  \BibitemOpen
  \bibfield  {author} {\bibinfo {author} {\bibfnamefont {J.~M.}\ \bibnamefont
  {Cowley}},\ }\href
  {https://doi.org/https://doi.org/10.1016/B978-0-444-82218-5.50025-0} {\emph
  {\bibinfo {title} {Diffraction Physics}}},\ \bibinfo {edition} {3rd}\ ed.,\
  North-Holland Personal Library\ (\bibinfo  {publisher} {Elsevier},\ \bibinfo
  {address} {Amsterdam},\ \bibinfo {year} {1995})\BibitemShut {NoStop}%
\bibitem [{\citenamefont {Lan}\ \emph {et~al.}(2015)\citenamefont {Lan},
  \citenamefont {Li}, \citenamefont {Hellman}, \citenamefont {Kim},
  \citenamefont {Mu{\~n}oz}, \citenamefont {Smith}, \citenamefont {Abernathy},\
  and\ \citenamefont {Fultz}}]{Lan2015}%
  \BibitemOpen
  \bibfield  {author} {\bibinfo {author} {\bibfnamefont {T.}~\bibnamefont
  {Lan}}, \bibinfo {author} {\bibfnamefont {C.~W.}\ \bibnamefont {Li}},
  \bibinfo {author} {\bibfnamefont {O.}~\bibnamefont {Hellman}}, \bibinfo
  {author} {\bibfnamefont {D.~S.}\ \bibnamefont {Kim}}, \bibinfo {author}
  {\bibfnamefont {J.~A.}\ \bibnamefont {Mu{\~n}oz}}, \bibinfo {author}
  {\bibfnamefont {H.}~\bibnamefont {Smith}}, \bibinfo {author} {\bibfnamefont
  {D.~L.}\ \bibnamefont {Abernathy}},\ and\ \bibinfo {author} {\bibfnamefont
  {B.}~\bibnamefont {Fultz}},\ }\href
  {https://doi.org/10.1103/PhysRevB.92.054304} {\bibfield  {journal} {\bibinfo
  {journal} {Phys. Rev. B}\ }\textbf {\bibinfo {volume} {92}},\ \bibinfo
  {pages} {054304} (\bibinfo {year} {2015})}\BibitemShut {NoStop}%
\bibitem [{\citenamefont {Dong}\ \emph {et~al.}(2000)\citenamefont {Dong},
  \citenamefont {Sankey}, \citenamefont {Ramachandran},\ and\ \citenamefont
  {McMillan}}]{Dong2000}%
  \BibitemOpen
  \bibfield  {author} {\bibinfo {author} {\bibfnamefont {J.}~\bibnamefont
  {Dong}}, \bibinfo {author} {\bibfnamefont {O.~F.}\ \bibnamefont {Sankey}},
  \bibinfo {author} {\bibfnamefont {G.~F.}\ \bibnamefont {Ramachandran}},\ and\
  \bibinfo {author} {\bibfnamefont {P.~F.}\ \bibnamefont {McMillan}},\ }\href
  {https://doi.org/10.1063/1.373447} {\bibfield  {journal} {\bibinfo  {journal}
  {J. Appl. Phys}\ }\textbf {\bibinfo {volume} {87}},\ \bibinfo {pages} {7726}
  (\bibinfo {year} {2000})}\BibitemShut {NoStop}%
\bibitem [{\citenamefont {Sergueev}\ \emph {et~al.}(2015)\citenamefont
  {Sergueev}, \citenamefont {Glazyrin}, \citenamefont {Kantor}, \citenamefont
  {McGuire}, \citenamefont {Chumakov}, \citenamefont {Klobes}, \citenamefont
  {Sales},\ and\ \citenamefont {Hermann}}]{Sergueev2015}%
  \BibitemOpen
  \bibfield  {author} {\bibinfo {author} {\bibfnamefont {I.}~\bibnamefont
  {Sergueev}}, \bibinfo {author} {\bibfnamefont {K.}~\bibnamefont {Glazyrin}},
  \bibinfo {author} {\bibfnamefont {I.}~\bibnamefont {Kantor}}, \bibinfo
  {author} {\bibfnamefont {M.~A.}\ \bibnamefont {McGuire}}, \bibinfo {author}
  {\bibfnamefont {A.~I.}\ \bibnamefont {Chumakov}}, \bibinfo {author}
  {\bibfnamefont {B.}~\bibnamefont {Klobes}}, \bibinfo {author} {\bibfnamefont
  {B.~C.}\ \bibnamefont {Sales}},\ and\ \bibinfo {author} {\bibfnamefont
  {R.~P.}\ \bibnamefont {Hermann}},\ }\href
  {https://doi.org/10.1103/PhysRevB.91.224304} {\bibfield  {journal} {\bibinfo
  {journal} {Phys. Rev. B}\ }\textbf {\bibinfo {volume} {91}},\ \bibinfo
  {pages} {224304} (\bibinfo {year} {2015})}\BibitemShut {NoStop}%
\bibitem [{\citenamefont {Ferreira}\ \emph {et~al.}(2020)\citenamefont
  {Ferreira}, \citenamefont {Paofai}, \citenamefont {L{\'e}toublon},
  \citenamefont {Ollivier}, \citenamefont {Raymond}, \citenamefont {Hehlen},
  \citenamefont {Ruffl{\'e}}, \citenamefont {Cordier}, \citenamefont {Katan},
  \citenamefont {Even},\ and\ \citenamefont {Bourges}}]{Ferreira2020}%
  \BibitemOpen
  \bibfield  {author} {\bibinfo {author} {\bibfnamefont {A.~C.}\ \bibnamefont
  {Ferreira}}, \bibinfo {author} {\bibfnamefont {S.}~\bibnamefont {Paofai}},
  \bibinfo {author} {\bibfnamefont {A.}~\bibnamefont {L{\'e}toublon}}, \bibinfo
  {author} {\bibfnamefont {J.}~\bibnamefont {Ollivier}}, \bibinfo {author}
  {\bibfnamefont {S.}~\bibnamefont {Raymond}}, \bibinfo {author} {\bibfnamefont
  {B.}~\bibnamefont {Hehlen}}, \bibinfo {author} {\bibfnamefont
  {B.}~\bibnamefont {Ruffl{\'e}}}, \bibinfo {author} {\bibfnamefont
  {S.}~\bibnamefont {Cordier}}, \bibinfo {author} {\bibfnamefont
  {C.}~\bibnamefont {Katan}}, \bibinfo {author} {\bibfnamefont
  {J.}~\bibnamefont {Even}},\ and\ \bibinfo {author} {\bibfnamefont
  {P.}~\bibnamefont {Bourges}},\ }\href
  {https://doi.org/10.1038/s42005-020-0313-7} {\bibfield  {journal} {\bibinfo
  {journal} {Commun. Phys.}\ }\textbf {\bibinfo {volume} {3}},\ \bibinfo
  {pages} {48} (\bibinfo {year} {2020})}\BibitemShut {NoStop}%
\bibitem [{\citenamefont {Barron}\ and\ \citenamefont
  {Klein}(1974)}]{BarronKlein1974}%
  \BibitemOpen
  \bibfield  {author} {\bibinfo {author} {\bibfnamefont {T.~H.~K.}\
  \bibnamefont {Barron}}\ and\ \bibinfo {author} {\bibfnamefont {M.~L.}\
  \bibnamefont {Klein}},\ }\bibinfo {title} {Perturbation theory of anharmonic
  crystals},\ in\ \href@noop {} {\emph {\bibinfo {booktitle} {Dynamical
  Properties of Solids}}},\ \bibinfo {editor} {edited by\ \bibinfo {editor}
  {\bibfnamefont {G.}~\bibnamefont {Horton}}\ and\ \bibinfo {editor}
  {\bibfnamefont {A.}~\bibnamefont {Maradudin}}}\ (\bibinfo  {publisher}
  {North-Holland},\ \bibinfo {address} {Amsterdam},\ \bibinfo {year} {1974})\
  p.\ \bibinfo {pages} {391}\BibitemShut {NoStop}%
\bibitem [{\citenamefont {Gardiner}\ and\ \citenamefont
  {Zoller}(2000)}]{Gardiner2000}%
  \BibitemOpen
  \bibfield  {author} {\bibinfo {author} {\bibfnamefont {C.~W.}\ \bibnamefont
  {Gardiner}}\ and\ \bibinfo {author} {\bibfnamefont {P.}~\bibnamefont
  {Zoller}},\ }\href@noop {} {\emph {\bibinfo {title} {Quantum Noise: A
  Handbook of Markovian and Non-Markovian Quantum Stochastic Methods with
  Applications to Quantum Optics}}}\ (\bibinfo  {publisher} {Springer},\
  \bibinfo {year} {2000})\BibitemShut {NoStop}%
\bibitem [{\citenamefont {Clerk}\ \emph {et~al.}(2010)\citenamefont {Clerk},
  \citenamefont {Devoret}, \citenamefont {Girvin}, \citenamefont {Marquardt},\
  and\ \citenamefont {Schoelkopf}}]{Clerk2010}%
  \BibitemOpen
  \bibfield  {author} {\bibinfo {author} {\bibfnamefont {A.~A.}\ \bibnamefont
  {Clerk}}, \bibinfo {author} {\bibfnamefont {M.~H.}\ \bibnamefont {Devoret}},
  \bibinfo {author} {\bibfnamefont {S.~M.}\ \bibnamefont {Girvin}}, \bibinfo
  {author} {\bibfnamefont {F.}~\bibnamefont {Marquardt}},\ and\ \bibinfo
  {author} {\bibfnamefont {R.~J.}\ \bibnamefont {Schoelkopf}},\ }\href
  {https://doi.org/10.1103/RevModPhys.82.1155} {\bibfield  {journal} {\bibinfo
  {journal} {Rev. Mod. Phys.}\ }\textbf {\bibinfo {volume} {82}},\ \bibinfo
  {pages} {1155} (\bibinfo {year} {2010})}\BibitemShut {NoStop}%
\end{thebibliography}%


\begin{thebibliography}{10}%
\makeatletter
\providecommand \@ifxundefined [1]{%
 \@ifx{#1\undefined}
}%
\providecommand \@ifnum [1]{%
 \ifnum #1\expandafter \@firstoftwo
 \else \expandafter \@secondoftwo
 \fi
}%
\providecommand \@ifx [1]{%
 \ifx #1\expandafter \@firstoftwo
 \else \expandafter \@secondoftwo
 \fi
}%
\providecommand \natexlab [1]{#1}%
\providecommand \enquote  [1]{``#1''}%
\providecommand \bibnamefont  [1]{#1}%
\providecommand \bibfnamefont [1]{#1}%
\providecommand \citenamefont [1]{#1}%
\providecommand \href@noop [0]{\@secondoftwo}%
\providecommand \href [0]{\begingroup \@sanitize@url \@href}%
\providecommand \@href[1]{\@@startlink{#1}\@@href}%
\providecommand \@@href[1]{\endgroup#1\@@endlink}%
\providecommand \@sanitize@url [0]{\catcode `\\12\catcode `\$12\catcode
  `\&12\catcode `\#12\catcode `\^12\catcode `\_12\catcode `\%12\relax}%
\providecommand \@@startlink[1]{}%
\providecommand \@@endlink[0]{}%
\providecommand \url  [0]{\begingroup\@sanitize@url \@url }%
\providecommand \@url [1]{\endgroup\@href {#1}{\urlprefix }}%
\providecommand \urlprefix  [0]{URL }%
\providecommand \Eprint [0]{\href }%
\providecommand \doibase [0]{https://doi.org/}%
\providecommand \selectlanguage [0]{\@gobble}%
\providecommand \bibinfo  [0]{\@secondoftwo}%
\providecommand \bibfield  [0]{\@secondoftwo}%
\providecommand \translation [1]{[#1]}%
\providecommand \BibitemOpen [0]{}%
\providecommand \bibitemStop [0]{}%
\providecommand \bibitemNoStop [0]{.\EOS\space}%
\providecommand \EOS [0]{\spacefactor3000\relax}%
\providecommand \BibitemShut  [1]{\csname bibitem#1\endcsname}%
\let\auto@bib@innerbib\@empty
\bibitem [{\citenamefont {Arnold}\ \emph {et~al.}(2014)\citenamefont {Arnold},
  \citenamefont {Bilheux}, \citenamefont {Borreguero}, \citenamefont {Buts},
  \citenamefont {Campbell}, \citenamefont {Chapon}, \citenamefont {Doucet},
  \citenamefont {Draper}, \citenamefont {{Ferraz Leal}}, \citenamefont {Gigg},
  \citenamefont {Lynch}, \citenamefont {Markvardsen}, \citenamefont
  {Mikkelson}, \citenamefont {Mikkelson}, \citenamefont {Miller}, \citenamefont
  {Palmen}, \citenamefont {Parker}, \citenamefont {Passos}, \citenamefont
  {Perring}, \citenamefont {Peterson}, \citenamefont {Ren}, \citenamefont
  {Reuter}, \citenamefont {Savici}, \citenamefont {Taylor}, \citenamefont
  {Taylor}, \citenamefont {Tolchenov}, \citenamefont {Zhou},\ and\
  \citenamefont {Zikovsky}}]{Arnold2013}%
  \BibitemOpen
  \bibfield  {author} {\bibinfo {author} {\bibfnamefont {O.}~\bibnamefont
  {Arnold}}, \bibinfo {author} {\bibfnamefont {J.}~\bibnamefont {Bilheux}},
  \bibinfo {author} {\bibfnamefont {J.}~\bibnamefont {Borreguero}}, \bibinfo
  {author} {\bibfnamefont {A.}~\bibnamefont {Buts}}, \bibinfo {author}
  {\bibfnamefont {S.}~\bibnamefont {Campbell}}, \bibinfo {author}
  {\bibfnamefont {L.}~\bibnamefont {Chapon}}, \bibinfo {author} {\bibfnamefont
  {M.}~\bibnamefont {Doucet}}, \bibinfo {author} {\bibfnamefont
  {N.}~\bibnamefont {Draper}}, \bibinfo {author} {\bibfnamefont
  {R.}~\bibnamefont {{Ferraz Leal}}}, \bibinfo {author} {\bibfnamefont
  {M.}~\bibnamefont {Gigg}}, \bibinfo {author} {\bibfnamefont {V.}~\bibnamefont
  {Lynch}}, \bibinfo {author} {\bibfnamefont {A.}~\bibnamefont {Markvardsen}},
  \bibinfo {author} {\bibfnamefont {D.}~\bibnamefont {Mikkelson}}, \bibinfo
  {author} {\bibfnamefont {R.}~\bibnamefont {Mikkelson}}, \bibinfo {author}
  {\bibfnamefont {R.}~\bibnamefont {Miller}}, \bibinfo {author} {\bibfnamefont
  {K.}~\bibnamefont {Palmen}}, \bibinfo {author} {\bibfnamefont
  {P.}~\bibnamefont {Parker}}, \bibinfo {author} {\bibfnamefont
  {G.}~\bibnamefont {Passos}}, \bibinfo {author} {\bibfnamefont
  {T.}~\bibnamefont {Perring}}, \bibinfo {author} {\bibfnamefont
  {P.}~\bibnamefont {Peterson}}, \bibinfo {author} {\bibfnamefont
  {S.}~\bibnamefont {Ren}}, \bibinfo {author} {\bibfnamefont {M.}~\bibnamefont
  {Reuter}}, \bibinfo {author} {\bibfnamefont {A.}~\bibnamefont {Savici}},
  \bibinfo {author} {\bibfnamefont {J.}~\bibnamefont {Taylor}}, \bibinfo
  {author} {\bibfnamefont {R.}~\bibnamefont {Taylor}}, \bibinfo {author}
  {\bibfnamefont {R.}~\bibnamefont {Tolchenov}}, \bibinfo {author}
  {\bibfnamefont {W.}~\bibnamefont {Zhou}},\ and\ \bibinfo {author}
  {\bibfnamefont {J.}~\bibnamefont {Zikovsky}},\ }\href
  {https://doi.org/https://doi.org/10.1016/j.nima.2014.07.029} {\bibfield
  {journal} {\bibinfo  {journal} {Nucl. Instrum. Meth. A}\ }\textbf {\bibinfo
  {volume} {764}},\ \bibinfo {pages} {156} (\bibinfo {year}
  {2014})}\BibitemShut {NoStop}%
\bibitem [{\citenamefont {Saunders}\ \emph {et~al.}(2022)\citenamefont
  {Saunders}, \citenamefont {Kim}, \citenamefont {Hellman}, \citenamefont
  {Smith}, \citenamefont {Weadock}, \citenamefont {Omelchenko}, \citenamefont
  {Granroth}, \citenamefont {Bernal-Choban}, \citenamefont {Lohaus},
  \citenamefont {Abernathy},\ and\ \citenamefont {Fultz}}]{Saunders2022}%
  \BibitemOpen
  \bibfield  {author} {\bibinfo {author} {\bibfnamefont {C.~N.}\ \bibnamefont
  {Saunders}}, \bibinfo {author} {\bibfnamefont {D.~S.}\ \bibnamefont {Kim}},
  \bibinfo {author} {\bibfnamefont {O.}~\bibnamefont {Hellman}}, \bibinfo
  {author} {\bibfnamefont {H.~L.}\ \bibnamefont {Smith}}, \bibinfo {author}
  {\bibfnamefont {N.~J.}\ \bibnamefont {Weadock}}, \bibinfo {author}
  {\bibfnamefont {S.~T.}\ \bibnamefont {Omelchenko}}, \bibinfo {author}
  {\bibfnamefont {G.~E.}\ \bibnamefont {Granroth}}, \bibinfo {author}
  {\bibfnamefont {C.~M.}\ \bibnamefont {Bernal-Choban}}, \bibinfo {author}
  {\bibfnamefont {S.~H.}\ \bibnamefont {Lohaus}}, \bibinfo {author}
  {\bibfnamefont {D.~L.}\ \bibnamefont {Abernathy}},\ and\ \bibinfo {author}
  {\bibfnamefont {B.}~\bibnamefont {Fultz}},\ }\href
  {https://doi.org/10.1103/PhysRevB.105.174308} {\bibfield  {journal} {\bibinfo
   {journal} {Phys. Rev. B}\ }\textbf {\bibinfo {volume} {105}},\ \bibinfo
  {pages} {174308} (\bibinfo {year} {2022})}\BibitemShut {NoStop}%
\bibitem [{\citenamefont {Hellman}\ and\ \citenamefont
  {Abrikosov}(2013)}]{Hellman2013}%
  \BibitemOpen
  \bibfield  {author} {\bibinfo {author} {\bibfnamefont {O.}~\bibnamefont
  {Hellman}}\ and\ \bibinfo {author} {\bibfnamefont {I.~A.}\ \bibnamefont
  {Abrikosov}},\ }\href {https://doi.org/10.1103/PhysRevB.88.144301} {\bibfield
   {journal} {\bibinfo  {journal} {Phys. Rev. B}\ }\textbf {\bibinfo {volume}
  {88}},\ \bibinfo {pages} {144301} (\bibinfo {year} {2013})}\BibitemShut
  {NoStop}%
\bibitem [{\citenamefont {Shapeev}(2016)}]{Shapeev2016}%
  \BibitemOpen
  \bibfield  {author} {\bibinfo {author} {\bibfnamefont {A.~V.}\ \bibnamefont
  {Shapeev}},\ }\href {https://doi.org/10.1137/15M1054183} {\bibfield
  {journal} {\bibinfo  {journal} {Multiscale Model Simul.}\ }\textbf {\bibinfo
  {volume} {14}},\ \bibinfo {pages} {1153} (\bibinfo {year}
  {2016})}\BibitemShut {NoStop}%
\bibitem [{\citenamefont {Ladygin}\ \emph {et~al.}(2020)\citenamefont
  {Ladygin}, \citenamefont {Korotaev}, \citenamefont {Yanilkin},\ and\
  \citenamefont {Shapeev}}]{Ladygin2020}%
  \BibitemOpen
  \bibfield  {author} {\bibinfo {author} {\bibfnamefont {V.~V.}\ \bibnamefont
  {Ladygin}}, \bibinfo {author} {\bibfnamefont {P.~Y.}\ \bibnamefont
  {Korotaev}}, \bibinfo {author} {\bibfnamefont {A.~V.}\ \bibnamefont
  {Yanilkin}},\ and\ \bibinfo {author} {\bibfnamefont {A.~V.}\ \bibnamefont
  {Shapeev}},\ }\href
  {https://doi.org/https://doi.org/10.1016/j.commatsci.2019.109333} {\bibfield
  {journal} {\bibinfo  {journal} {Computational Materials Science}\ }\textbf
  {\bibinfo {volume} {172}},\ \bibinfo {pages} {109333} (\bibinfo {year}
  {2020})}\BibitemShut {NoStop}%
\bibitem [{\citenamefont {Novikov}\ \emph {et~al.}(2020)\citenamefont
  {Novikov}, \citenamefont {Gubaev}, \citenamefont {Podryabinkin},\ and\
  \citenamefont {Shapeev}}]{Novikov2020}%
  \BibitemOpen
  \bibfield  {author} {\bibinfo {author} {\bibfnamefont {I.~S.}\ \bibnamefont
  {Novikov}}, \bibinfo {author} {\bibfnamefont {K.}~\bibnamefont {Gubaev}},
  \bibinfo {author} {\bibfnamefont {E.~V.}\ \bibnamefont {Podryabinkin}},\ and\
  \bibinfo {author} {\bibfnamefont {A.~V.}\ \bibnamefont {Shapeev}},\ }\href
  {https://doi.org/10.1088/2632-2153/abc9fe} {\bibfield  {journal} {\bibinfo
  {journal} {Mach. Learn.: Sci. Technol.}\ }\textbf {\bibinfo {volume} {2}},\
  \bibinfo {pages} {025002} (\bibinfo {year} {2020})}\BibitemShut {NoStop}%
\bibitem [{\citenamefont {Kresse}\ and\ \citenamefont
  {Hafner}(1993)}]{Kresse1993}%
  \BibitemOpen
  \bibfield  {author} {\bibinfo {author} {\bibfnamefont {G.}~\bibnamefont
  {Kresse}}\ and\ \bibinfo {author} {\bibfnamefont {J.}~\bibnamefont
  {Hafner}},\ }\href {https://doi.org/10.1103/PhysRevB.47.558} {\bibfield
  {journal} {\bibinfo  {journal} {Phys. Rev. B}\ }\textbf {\bibinfo {volume}
  {47}},\ \bibinfo {pages} {558} (\bibinfo {year} {1993})}\BibitemShut
  {NoStop}%
\bibitem [{\citenamefont {Kresse}\ and\ \citenamefont
  {Hafner}(1994)}]{Kresse1994}%
  \BibitemOpen
  \bibfield  {author} {\bibinfo {author} {\bibfnamefont {G.}~\bibnamefont
  {Kresse}}\ and\ \bibinfo {author} {\bibfnamefont {J.}~\bibnamefont
  {Hafner}},\ }\href {https://doi.org/10.1103/PhysRevB.49.14251} {\bibfield
  {journal} {\bibinfo  {journal} {Phys. Rev. B}\ }\textbf {\bibinfo {volume}
  {49}},\ \bibinfo {pages} {14251} (\bibinfo {year} {1994})}\BibitemShut
  {NoStop}%
\bibitem [{\citenamefont {Kresse}\ and\ \citenamefont
  {Furthm{\"u}ller}(1996)}]{Kresse1996_1}%
  \BibitemOpen
  \bibfield  {author} {\bibinfo {author} {\bibfnamefont {G.}~\bibnamefont
  {Kresse}}\ and\ \bibinfo {author} {\bibfnamefont {J.}~\bibnamefont
  {Furthm{\"u}ller}},\ }\href {https://doi.org/10.1016/0927-0256(96)00008-0}
  {\bibfield  {journal} {\bibinfo  {journal} {Comput. Mater. Sci.}\ }\textbf
  {\bibinfo {volume} {6}},\ \bibinfo {pages} {15 } (\bibinfo {year}
  {1996})}\BibitemShut {NoStop}%
\bibitem [{\citenamefont {Kresse}\ and\ \citenamefont
  {Furthm\"uller}(1996)}]{Kresse1996_2}%
  \BibitemOpen
  \bibfield  {author} {\bibinfo {author} {\bibfnamefont {G.}~\bibnamefont
  {Kresse}}\ and\ \bibinfo {author} {\bibfnamefont {J.}~\bibnamefont
  {Furthm\"uller}},\ }\href {https://doi.org/10.1103/PhysRevB.54.11169}
  {\bibfield  {journal} {\bibinfo  {journal} {Phys. Rev. B}\ }\textbf {\bibinfo
  {volume} {54}},\ \bibinfo {pages} {11169} (\bibinfo {year}
  {1996})}\BibitemShut {NoStop}%
\end{thebibliography}%

\end{document}


\preprint{APS/123-QED}

\title{Supplemental Material for Diffuse Inelastic Neutron Scattering from Anharmonic Vibrations in Cuprite}
\author{C. N. Saunders$^{\S}$}
\email{clairenicolesaunders@gmail.com}
\author{V. V. Ladygin$^\S$}
\email{vladygin@caltech.edu}
\affiliation{California Institute of Technology, Pasadena, California 91125, USA}
\author{D. S. Kim}
\affiliation{The University of California, Los Angeles, Los Angeles, California 90095, USA}
\author{C. M. Bernal-Choban}
\author{S. H. Lohaus}
\affiliation{California Institute of Technology, Pasadena, California 91125, USA}
\author{G. E. Granroth}
\author{D. L. Abernathy}
\affiliation{Oak Ridge National Laboratory, Oak Ridge, Tennessee 37831, USA}
\author{B. Fultz}
\email{btf@caltech.edu}
\affiliation{California Institute of Technology, Pasadena, California 91125, USA}

\date{\today}
\maketitle
\def\thefootnote{$\S$}\footnotetext{Equally contributing authors.}

\section{Single Crystal Data Analysis}
We used the software Mantid to reduce the single crystal data to the four-dimensional \textit{S}($\vec{Q}$,$\varepsilon$) \cite{Arnold2013}. An additional analysis assessed the data statistics, alignment, nonlinearities, and possible abnormalities in the different Brillouin zones. We found no problems, so using symmetry operators, we folded data throughout $\vec{Q}$ into an irreducible wedge in the first Brillouin zone. Before folding each zone, we subtracted an averaged incoherent multiphonon scattering correction. 

After folding, we thermally-weighted the spectral intensities for better visualization of the high energy modes using the factor:
\begin{equation}
\varepsilon \left[1 - \mathrm{exp}\left(\frac{-\varepsilon}{k_{B}T}\right)\right]\times\textit{S}(\vec{Q},\varepsilon)
 = \textit{S}_{\mathrm{weighted}}(\vec{Q},\varepsilon)
\end{equation}
where $\varepsilon$ = $\hbar\omega$. 
More explanation of this data processing appears in the Supplemental of our prior publication on cuprite \cite{Saunders2022}.

\section{stochastic Temperature-Dependent Effective Potential Method}

We calculated cubic anharmonic corrections to the phonon self-energy with the stochastic Temperature-Dependent Effective Potential (sTDEP) package \cite{Hellman2013}. The corrections to the phonon self-energy give phonon spectra with thermal shifts and finite linewidth. 

By solving a dynamical matrix, we obtained phonon frequencies. For a given third-order force constant, $\Phi_{ss^{'}s^{''}}$, we calculated and corrected the phonon self-energy with the real, $\Delta$, and imaginary, $\Gamma$, corrections to the phonon self-energy. The imaginary correction is:
\begin{widetext}
\begin{align}
\label{eq:calc7}
\Gamma_{s}^{\vec{q}}(V,T) &= \frac{\hbar\pi}{16}\sum_{\vec{q}^{\prime},s^{\prime},\vec{q}^{\prime\prime},s^{\prime\prime}}\left|\Phi_{ss^{'}s^{\prime\prime}}^{\vec{q}\vec{q}^{\prime}\vec{q}^{\prime\prime}}\right|^2  \left(n_{\vec{q}^{\prime},s^{\prime}}+n_{\vec{q}^{\prime\prime},s^{\prime\prime}}+1\right) 
\times\delta\left(\Omega-\omega_{\vec{q}^{\prime},s^{\prime}}-\omega_{\vec{q}^{\prime\prime},s^{\prime\prime}}\right) 
\\
&+ \left(n_{\vec{q^{\prime}},s^{\prime}}-n_{\vec{q^{\prime\prime}},s^{\prime\prime}}\right)\left[\delta\left(\Omega-\omega_{\vec{q^{\prime}},s^{\prime}}+\omega_{\vec{q^{\prime\prime}},s^{\prime\prime}}\right)-\delta\left(\Omega+\omega_{\vec{q^{\prime}},s^{\prime}}-\omega_{\vec{q^{\prime\prime}},s^{\prime\prime}}\right)\right] 
\end{align}
\end{widetext}
where $\hbar\Omega$ is a probing energy, $\omega_{\vec{q},s}^2$ are the eigenvalues of the dynamical matrix, and $n$ are the Planck occupancy factors. The three-phonon matrix component is:
\begin{widetext}
\begin{align}
\Phi_{ss^{\prime}s^{\prime\prime}}^{\vec{q}\vec{q}^{\prime}\vec{q}^{\prime\prime}}&=\sum_{ijk}\sum_{\alpha\beta\gamma}\frac{\epsilon_{s}^{i\alpha}\epsilon_{s^{\prime}}^{j\beta}\epsilon_{s^{\prime\prime}}^{k\gamma}}{\sqrt{m_i m_j m_k}\sqrt{\omega_{\vec{q},s}\omega_{\vec{q}^{\prime},s^{\prime}}\omega_{\vec{q}^{\prime\prime},s^{\prime\prime}}}}\Phi_{ijk}^{\alpha\beta\gamma}\mathrm{exp}\left[{i\left(\vec{q}\cdot\vec{r_i}+\vec{q}^{\prime}\cdot\vec{r_j}+\vec{q}^{\prime\prime}\cdot\vec{r_k}\right)}\right]
\end{align}
\end{widetext}
where the primes identify the three phonons. 

Finally, we obtain the real part of the phonon self-energy correction from the Kramers--Kronig transform:
\begin{equation}
\label{eq:calc9}
    \Delta_{\vec{q},s}(\Omega_{\vec{q},s}) = \frac{1}{\pi}\int\frac{\Gamma(\omega_{\vec{q},s})}{\omega_{\vec{q},s}-\Omega}d\omega_{\vec{q},s}
\end{equation}
Large deviations of $\Delta_{\vec{q},s}(\Omega_{\vec{q},s})$ from a Lorentzian function suggest a high degree of anharmonicity. 

\section{Higher-Order Anharmonicity and Coherence}
Our previous report on the thermal expansion of cuprite near room temperature \cite{Saunders2022} describes in detail the sTDEP computations used for the results in Fig. \ref{fig:TDEPresultsSM} below. This earlier work suggested that phonon interactions beyond the third order are required, even at \qty{300}{\kelvin}. A comparison of Fig. 1 to the experimental dispersions in the main text shows that the linewidths from sTDEP are qualitatively inadequate to account for the phonon spectra of cuprite at \qty{700}{\kelvin} and \qty{900}{\kelvin}. The phonon self-energy corrections for the fourth-order loop diagram do not have an imaginary component, so  better calculations with a many-body theory approach may require five-phonon processes or higher. The phase space for such multi-phonon kinematics is vast. To our knowledge, there are no reports of complete calculations with five-phonon processes. 
Higher-order perturbation theory may not be appropriate for calculations of DII.

In principle, the MLIP calculations include effects from higher-order phonon anharmonicity, although it was not practical to isolate individual contributions. With the MLIP approach, there is a significant improvement in the agreement with the experimental results at \qty{700}{\kelvin} and \qty{900}{\kelvin} when the Langevin thermostat is employed. Our interpretation of the DII requires randomness in the anharmonic interactions, not just large anharmonicity. The Langevin thermostat offers more accurate simulations, but does not account fully for the time dynamics of the O-atoms in cuprite. 

The broad energy spread of the experimental inelastic intensity at high temperatures suggests that atom vibrations lose coherence quickly. The DII has characteristics of diffuse scattering from displacement disorder in elastic scattering, although the variables for DII are energy and time instead of momentum and position for elastic scattering. 

\begin{figure}
\centering
\includegraphics[width=0.85\columnwidth]{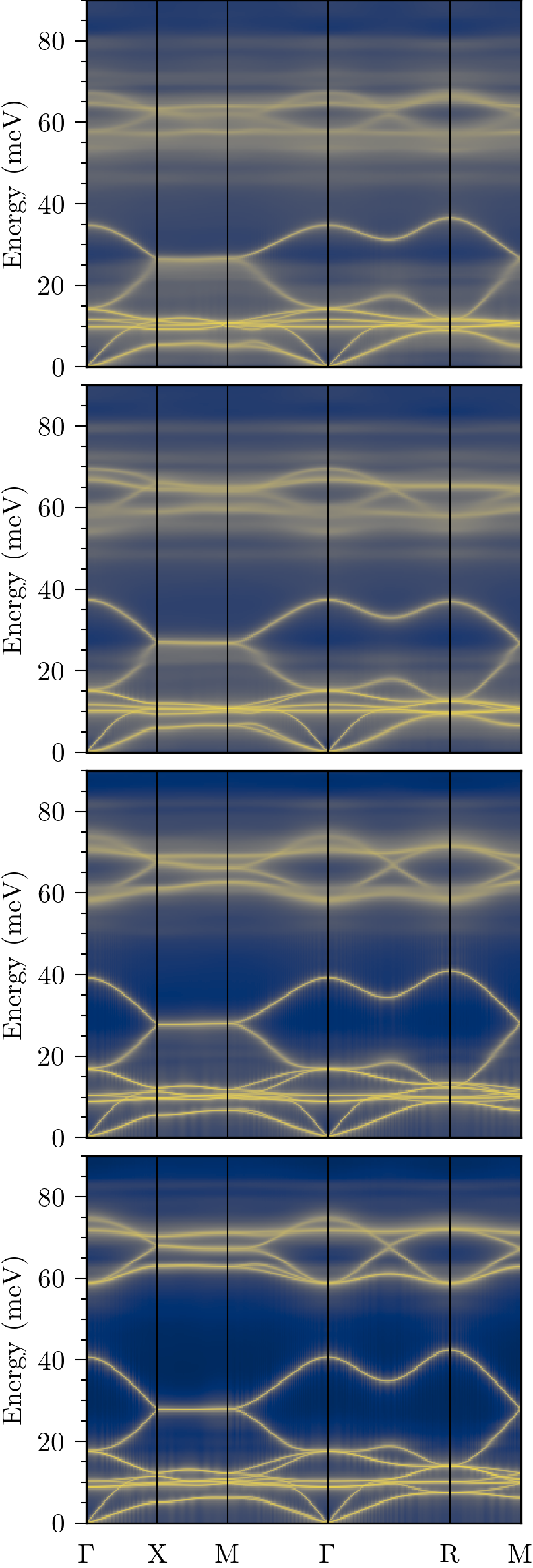}
\caption{Calculated sTDEP phonon dispersions along high-symmetry directions for temperatures from top to bottom: \qty{900}{\kelvin}, \qty{700}{\kelvin}, \qty{300}{\kelvin}, and \qty{10}{\kelvin}.}
\label{fig:TDEPresultsSM}
\end{figure}

\section{Machine Learning Interatomic Potentials (MLIP) Construction}

We used moment tensor potentials (MTP) \cite{Shapeev2016} to model the interatomic interactions. Moment tensors are flexible basis functions helpful in learning the lattice dynamics of a system, including anharmonic effects \cite{Ladygin2020}. The construction of machine learning interatomic potentials (MLIP) followed the procedure of Ref. \cite{Novikov2020}. We generated the MLIP from configurations actively sampled from molecular dynamics simulations at temperatures from 10\,K to 900\,K. The training set contained 523 configurations post-processed with VASP providing reference energies, forces, and stresses \cite{Kresse1993, Kresse1994, Kresse1996_1, Kresse1996_2}. The average difference between the fitted potential and the potential from the \textit{ab initio} model was \qty{0.4}{\meV/atom}. Using Eq. \ref{eq:forcerror}, we calculated a relative difference of 5~\% in forces, and 4~\% in stresses. In detail, 
\begin{equation}
A_{rel} = \sqrt{\frac {\langle (\Delta A - \overline{\Delta A})^2 \rangle} {\langle (A_{MTP} - \overline{A}_{MTP})^2\rangle}},
    \label{eq:forcerror}
\end{equation}
where $\Delta A$ is the difference in force or stress between \textit{ab initio} model and moment tensor potential, and $\overline{\Delta A}$ its mean value. Quantities from the moment tensor potentials are designated by subscript $MTP$. 

\section{Projected velocity-velocity correlation functions}
Correlation functions provide statistical relationships between random variables. The correlation $C(x,t)$ of the random variable $A(x)$, a function of space, and $B(t)$, a random function of time is 
\begin{equation}
    C(x,t)=\langle A(x)B(t)\rangle,
    \label{eq:correlation1}
\end{equation}
where the brackets represent the expectation value of the variables. For  correlations of random independent variables of vector functions, Eq. \ref{eq:correlation1} is modified for vector components
\begin{equation}
    C_{ij}(x,t)=\langle A_i(x) B_j(t)\rangle
    \label{eq:correlation2a}
\end{equation}
Here $C_{ij}(x,t)$ is the ${ij}^{th}$ component of the correlation matrix, written as
\begin{equation}
C(x,t)=\sum_i\sum_j\langle A_i(x) B_j(t)\rangle.   \label{eq:correlation2b}
\end{equation}
Studies of lattice dynamics require time and space relationships. Autocorrelation functions give correlations of the same quantity at two distinct points. For random variables $A(x)$ and $B(t)$ 
\begin{equation}
    C_{AA}(x_1,x_2)=\langle A(x_1) A(x_2)\rangle
    \label{eq:CAA}
\end{equation}
and
\begin{equation}
    C_{BB}(t_1,t_2)=\langle B(t_1) B(t_2)\rangle,
    \label{eq:CBB}
\end{equation}
for the spatial autocorrelation of $A$, $C_{AA}$, and the temporal autocorrelation of $B$, $C_{BB}$. 

For phonon dispersions, we want the autocorrelation of the position of an atom at two different times. 
We evaluated the correlation with the mode-projected velocity, a form of current density
\begin{equation}
    \vec{j}(\vec{q}, t) = \sum_i^{N} \vec{v}_i (t)\,\mathrm{exp}[{i \vec{q}\cdot\vec{r}_i(t)}].
    \label{eq:j_A}
\end{equation}
Here $\vec{q}$, $\vec{r}_i(t)$, and $\vec{v}_i(t)$ are the respective wavevector, position, and mass-weighted velocity of atom $i$ at time $t$ with the current density separated into longitudinal ($L$) and transverse ($T$) components to the direction of the wavevector.
\begin{equation}
\begin{split}
\vec{j}_L(\vec{q}, t) &= \hat{q} \,\sum_i^{N} (\vec{v}_i(t) \cdot \hat{{q}})\,\mathrm{exp}[{i\vec{q}\cdot\vec{r}_i(t)}]\\
\vec{j}_T(\vec{q}, t) &= \sum_i^{N} \left[\vec{v}_i(t) - \hat{q} (\vec{v}_i(t) \cdot \hat{q}) \right] \mathrm{exp}[{i \vec{q}\cdot\vec{r}_i(t)}].
\end{split}
\end{equation}

The Fourier transforms of the longitudinal and transverse  autocorrelation function $\Big\langle\vec{j}_{\kappa}(t, \vec{q})\vec{j}_{k}(0, -\vec{q})\Big\rangle$ give a partial vibrational density of states at a particular point of the Brillouin zone,
\begin{equation}
    g_{\kappa}(\omega, \vec{q}) = \int_0^{\infty} \Big\langle\vec{j}_{\kappa}(t, \vec{q})\vec{j}_{\kappa}(0, -\vec{q})\Big\rangle \,\mathrm{exp}({-i\omega t})\, dt,
\end{equation}
where $\kappa$ denotes either $L$ or $T$.
Unlike perturbation theory methods, autocorrelation methods contain all orders of anharmonicity. However, autocorrelation functions require substantially larger time and length scales to reach convergence than methods based on perturbation theory.

\section{Models for Diffuse Inelastic Intensity Spectra}

We expect a semi-periodic structure in the fluctuation function $l(t)$ as O-atoms vibrate against Cu neighbors. Figure 3 in Ref. \cite{Saunders2022} shows the partial phonon DOS of Cu-atoms and for O-atoms of cuprite at 300\,K, indicating that we can approximate the partial DOS of Cu-atoms as a mode with energy 10\,meV, and the partial DOS of  O-atoms as a mode at approximately 72\,meV. With these very different frequencies, the dynamics of Cu- and O-atoms can be considered independent Einstein modes. Nevertheless, the O-atom does not behave like a simple harmonic oscillator for more than a few cycles. Instead, its tetrahedral cage of Cu-atoms undergoes deformations that alter the phase and amplitude of the O-atom dynamics. With a strongly nonlinear potential, this interaction disrupts the harmonic dynamics at close approaches of Cu- and O-atoms, giving a sudden change in force on the O-atom. Although this alters the number of phonons in the O-atom vibrational mode, over time, this mode recovers an average energy of $k_{\mathrm{B}}T$. However, there is a change in the phase of the oscillation of the O-atom that is cumulative over time.

Here is a more detailed model for instantaneous changes in the phase of O-atom vibrations. Unlike the model in the main text, there is no damped harmonic oscillator representing a phonon mode. Here $P_l(t)$ includes the periodic vibrations. We assume a Gaussian distribution of phase changes for each anharmonic interaction with a Cu-atom. The changes in phase average to zero but have a mean-squared error that grows with time. In our model, the characteristic time $\tau$ for the phase errors of the O-atom is the period of oscillation of the Cu-atom  ($\tau = 2 \pi / \omega_{\mathrm{Cu}} $). Each stochastic change of $\psi_{\rm O}(t)$ is assumed to have a Gaussian probability distribution with zero mean and a standard deviation, $\gamma$, so for the first interaction near the time $\tau$
\begin{eqnarray}
     P_{1}(t) =\frac{1}{\sqrt{\pi} \, \gamma} \, {\mathrm{exp}}[{-(t-\tau)^{2} / \gamma^{2}}] \; . 
\label{eq. 9.92SM}
\end{eqnarray}
The second interaction with the Cu-atom adds to the phase uncertainty of the O-atom. With respect to the phase at the initial time, it is the convolution of Eq. \ref{eq. 9.92SM} with itself, giving a Gaussian with standard deviation $\sqrt{2}{\gamma}$. The convolution with Eq. (\ref{eq. 9.92SM}) is performed at each time interval $\tau$, 
\begin{eqnarray}
     P_{l}(t)=\sum_{n=-\infty}^{\infty}\frac{1}{\sqrt{\left|n\right|\pi}\, \gamma} \, {\mathrm{exp}}[{-(t-n\tau)^{2}/(\left|n\right|
\gamma^{2})}]  \; 
\label{eq. 9.99SM}
\end{eqnarray}
which is a sum of Gaussians spaced by intervals of $\tau$, with widths increasing with time. The case $n=0$ is the limit $\tau \rightarrow 0$, giving a $\delta$-function of unit area. (Terms with negative $n$ give the  time distributions of interactions that give the reference phase at $t=0$.)

The diffuse inelastic intensity from the fluctuation term of Eq. 9 in the main text 
is
\begin{eqnarray}
I(\omega)  &=& {\mathcal K} \int_{-\infty}^{\infty} \mathrm{exp}({{ i} \omega t}) \,
V^2 P_l(t)
  \, {\rm d}t \; , \label{ScatteredIntensity2SMa}  \\
\frac{I(\omega)} {V^2 \mathcal K} &=& \int_{-\infty}^{\infty} \mathrm{exp}({{ i} \omega t}) \,
P_l(t)
  \, {\rm d}t \; , \label{ScatteredIntensity2SM}  \\
    \frac{I(\omega)} {V^2 \mathcal K}  &=&  
     \int\limits_{-\infty}^{\infty} \mathrm{exp}({{ -i} \omega t})  \\
     &\times&\! \sum_{n=-\infty}^{\infty}  \! \frac{1}
     {\sqrt{\left|  n\right|  \pi} \, \gamma}{\mathrm{exp}}[{-(t-n\tau)^{2}/(\left|n\right|
\gamma^{2})}] \, {\rm d} t \; \nonumber.
\label{Fourier_of_GaussiansSM}
\end{eqnarray}
Substituting $t^{\prime}=t-n \tau$ and taking Fourier transforms
of the series of Gaussian functions:
\begin{align}
    \frac{I(\omega)} {V^2 \mathcal K} =\sum_{n=-\infty}^{\infty} \! \mathrm{exp}({- \omega^{2}|n|\gamma^{2}
    /4}) \ \mathrm{exp}({-{i} n \omega \tau}) \; .
    \label{eq. 9.100SM}
\end{align}
Evaluating Eq. (\ref{eq. 9.100SM}) as two geometric series 
for positive and negative $n$
\begin{align}
    \frac{I(\omega)} {V^2 \mathcal K} &= \frac
    {1}{1-\exp\big[-\omega^{2}\gamma^{2}/4 + {{i}} \omega \tau \big]} + c.c. -1\; , 
    \label{eq. 9.105SM} \\
   \frac{I(\omega)} {V^2 \mathcal K} &=\frac{1-{\mathrm{exp}}[{-\omega^{2}\gamma^{2}/2}]}{1+\mathrm{exp}({- \omega^{2}
    \gamma^{2}/2}-2)\,\mathrm{exp}[{-\omega^{2}\gamma^{2}/4}]\cos( \omega \tau )} \; .
    \label{eq. 9.106SM}
\end{align}

\begin{figure}[ht]
\includegraphics[width=0.6\textwidth]{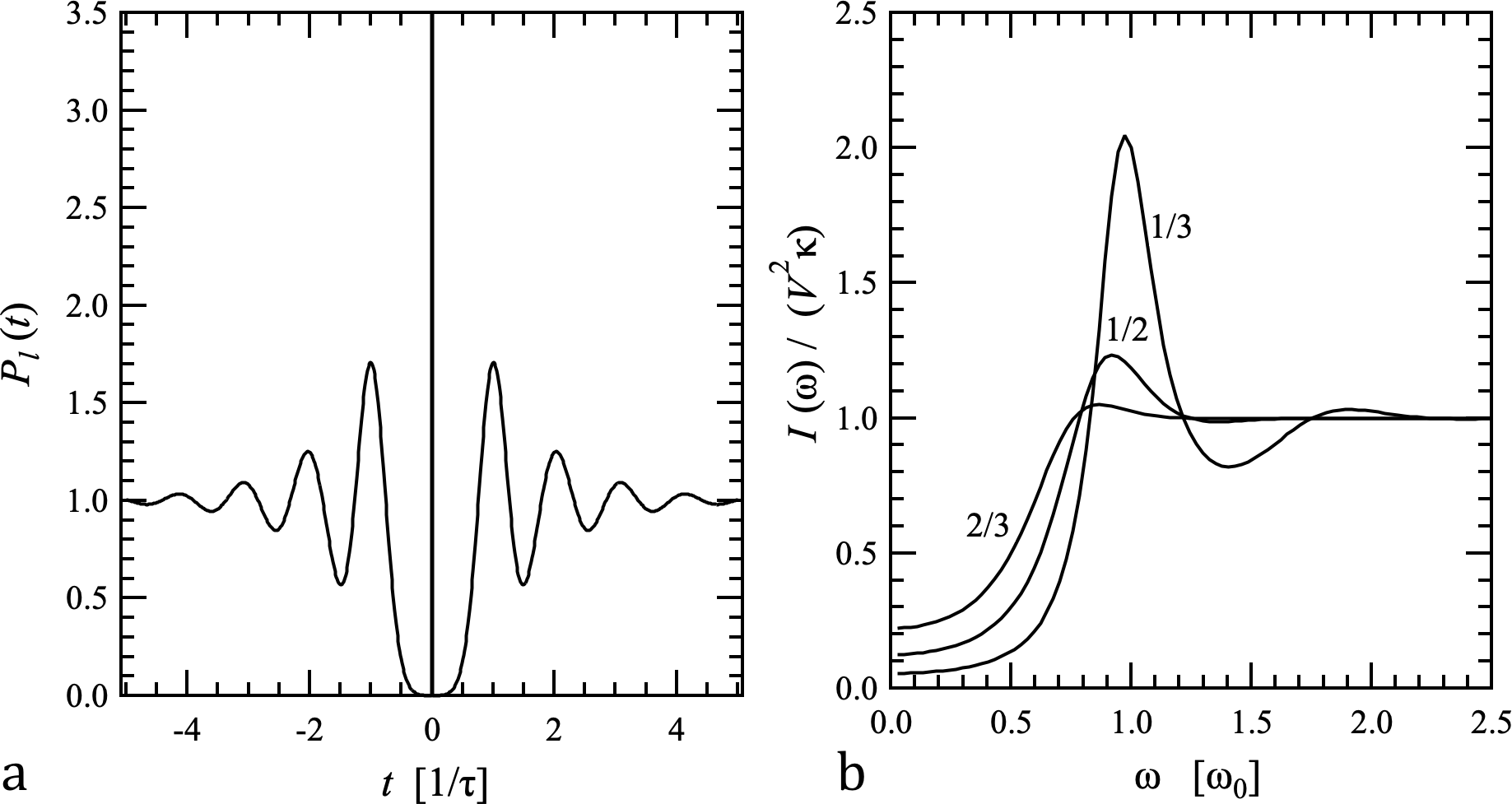}
\caption{({a}) Time-time correlation function of Eq. (\ref{eq. 9.99SM}) for  $\gamma= \tau/3$.
({b}) Intensity $I(\omega)$ of Eq. \ref{eq. 9.106SM} from Fourier transform of panel a, for different ratios of $\gamma / \tau$.}
\label{PhaseNoiseModelFunctionsSM}
\end{figure}

A time-time correlation function $P_l(t)$ is shown in Fig. \ref{PhaseNoiseModelFunctionsSM}a for $\gamma = \tau/3$. The corresponding $I(\omega)$ of Eq. (\ref{eq. 9.106SM}) is shown in Fig. \ref{PhaseNoiseModelFunctionsSM}b for this $\gamma = \tau/3$ and two other values. The fluctuation spectrum is highly sensitive to the ratio $\gamma/ \tau$. Small values of $\gamma/ \tau$  give a sharp peak in the inelastic spectrum at the expected vibrational frequency $\omega_0 = 2 \pi / \tau$.  
For many values of $\gamma$, at low $\omega$ the $I(\omega)$ of Eq. \ref{eq. 9.106SM} 
rises with $\omega$, and has a plateau at approximately 2$\omega_{0}$. The roll-off at low frequencies 
in Fig. \ref{PhaseNoiseModelFunctionsSM} seems consistent with the experimental data in Fig. 1 of the main text.

An assumption so far is that the noise term has only an instantaneous time correlation, so $P_l(t)$ has a $\delta$-function at $t=0$ in Fig. \ref{PhaseNoiseModelFunctionsSM}a. An instantaneous time correlation of  $P_l(t) = \delta(t)$ requires an instantaneous change in the positions of neighboring atoms responsible for $l(t)$. Classically, this requires an infinite force, or quantum mechanically, the energy operator ${\rm i}\hbar \, \partial \psi / \partial {t}$ is infinite. 

Physically, we expect interatomic anharmonic forces to be comparable to the forces on vibrating atoms, so the time correlation of $P_l(t)$ should be some fraction of the atom's vibrational period. A function with some width in time should be convoluted with the $P_l(t)$ above, and the $\delta$ function for $t=0$ will be broadened. This convolution is equivalent to multiplying the $I(\omega)$ in Fig. \ref{PhaseNoiseModelFunctionsSM}b by a factor such as $\widetilde{\mathcal P}_{{\Delta \omega}}(\omega)$, presented in Eq. 9 of the main text. So although the $I(\omega)$ from instantaneous interatomic interactions extends to infinity in $\omega$, the finite time of the interaction provides attenuation at large $\omega$. It may be possible to determine a characteristic time of anharmonic interatomic interactions by experimental measurements of the width of the DII. 

Two new parameters in the previous discussion are $V$ and $\gamma$. The parameter $V$ gives the coupling amplitude between a vibrational mode and the surrounding thermal bath. It plays the role of $\Phi({\vec{k}_1, \vec{k}_2, \vec{k}_3})$ in three-phonon processes, but $V$ ignores the details of the wavevectors and the kinematics of energy and momentum conservation. On the other hand, the parameter $V$ includes the phonon occupancy numbers, so $V$ is expected to increase with $T$. In classical dynamics, the parameter $\gamma$ gives the distribution of the changes of phase of an O-atom for each non-harmonic interaction with a Cu-atom. The parameter $\gamma$ value should also increase with larger thermal displacements at larger $T$. 


\onecolumngrid\

\section{Analyses of characteristic damping time}

\begin{figure}
\centering
\includegraphics[width=0.9\textwidth]{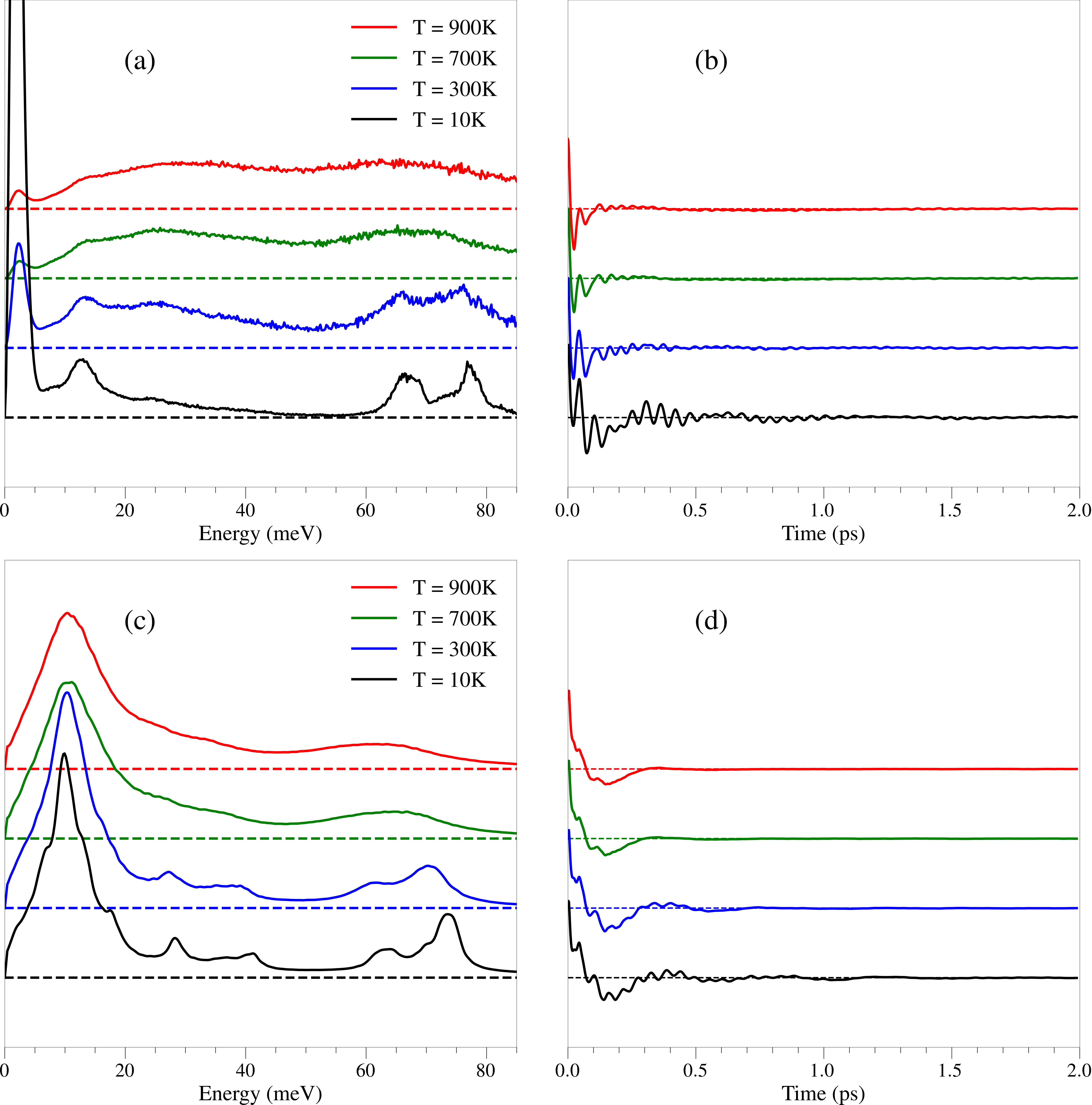}
\caption{Average phonon density of states from Fig. 1 in the main text (left) and its Fourier transform (right). (a), (b) for experimental data, and (c), (d)  for molecular dynamics simulations with a Langevin thermostat.  Temperatures from top to bottom are: \qty{900}{\kelvin}, \qty{700}{\kelvin}, \qty{300}{\kelvin}, and \qty{10}{\kelvin}.}
\label{fig:FourierAnalyses}
\end{figure}


To estimate the characteristic damping time of vibrations in cuprite, we first obtained a phonon density of states (DOS) for the experimental data, and for molecular dynamics simulations with the Langevin thermostat, using the data that provided Fig. 1 of the main text. We then performed  Fourier transforms of these phonon DOS. The results of this analyses are presented in Fig. \ref{fig:FourierAnalyses}. From Fig. \ref{fig:FourierAnalyses}a and c, we can estimate the characteristic decay time of the autocorrelation function (Fourier transform of PDOS) which is approximately equal \qty{3e-13}{\second} ($\gamma / \tau_O$= 6) at \qty{300}{\kelvin} to \qty{1e-13}{\second} ($\gamma / \tau_O$= 2) at \qty{900}{\kelvin}. These are consistent with the estimates in the main text.

\bibliography{supplemental.bib}